\documentclass[%
 reprint,
 amsmath,amssymb,
 aps,
]{revtex4-2}

\usepackage{dcolumn}
\usepackage{bm}

\usepackage{graphicx}
\usepackage{subfigure}
\usepackage{graphicx,hyperref}
\expandafter\let\csname equation*\endcsname\relax
\expandafter\let\csname endequation*\endcsname\relax
\usepackage{amsmath}
\usepackage{amssymb}
\usepackage{multirow, makecell, comment} 
\newcommand{\fla}[1]{\begin{flalign}#1\end{flalign}}
\usepackage{braket}
\usepackage{lineno}
\usepackage{lipsum}
\usepackage{xcolor}
\usepackage{soul}
\usepackage{array} 
\def\abs#1{|\,#1\,|}

\begin{document}

\preprint{APS/123-QED}

\title{Multiresonator quantum memory with atomic ensembles}


\author{  S.A. Moiseev}
\email{samoi@yandex.ru}

\affiliation{Kazan Quantum Center, Kazan National Research Technical University n.a. A.N. Tupolev-KAI, 10 K. Marx St., 420111, Kazan, Russia}

\date{\today}

\begin{abstract}

The theory of multiresonator quantum memory with atomic ensembles has been developed.  Using  the obtained analytical solutions, the basic physical properties of such memory are analyzed and optimal conditions for its implementation are determined. 
Advantages of this quantum memory and its experimental implementation in integrated optical  schemes are discussed.

\end{abstract}

                              
\maketitle

\section{Introduction}

The use of atomic ensembles    
\cite{Lukin2003,Lvovsky-NatPhot-2009, Hammerer2010,Chaneliere2018,Guo2023,Lei2023,2024-Moiseev-Physics-Uspekhi}
is one of the most promising approaches 
in creating multi-qubit quantum memory (QM) \cite{Azuma2023,Ye2024}.
In the last decade, the optical QM schemes have been actively developed, in which atomic ensembles are placed in optical resonators \cite{Sabooni2013,2021-PRB-Minnegaliev,Zhou_Review_2023,Duranti2024,Kollath_2025,Liu2025,Lau2025}.
The interest in developing resonator QM schemes is due to a number of interesting properties and advantages.
The use of resonators enhances the interaction of photons with atoms, which makes it possible to reduce the optical density of the resonant transition of the atomic ensemble, thereby significantly facilitating the conditions for achieving highly efficient QM.
The latter, in turn, facilitates the achievement of a longer coherence time of resonant transitions due to the significant suppression of interparticle interactions.
Furthermore, the QM cell size can be reduced in this way. Compact resonators can also be directly integrated into waveguide circuits, opening the possibility of creating compact quantum devices that incorporate QM along with various processors.

The greatest advantages of resonator QM are achieved by using high-quality resonators, which allow maximizing the interaction of photons with atoms.
However, placing an atomic ensemble in a single resonator connected to an external waveguide with a constant coupling $\kappa$ cannot provide effective memory in a spectral width greater than $\kappa/3$ \cite{Moiseev2010cavity}.
Accordingly, increasing the resonator's Q factor
narrows the operating spectral range of the QM.
Expansion of this spectral range is possible through the use of so-called white resonators \cite{EMoiseev2021}  and by combining several resonators with different frequencies into a common circuit connected to the waveguide directly \cite{EMoiseev2016}, or through a common resonator \cite{Moiseev2018}.
The multiresonator schemes themselves have already proven promising for microwave QM  \cite{2021-Bao-PRL}, and have made it possible to achieve a fairly high efficiency of $\sim 60 \%$  when storing Gaussian single-photon wave packets \cite{Matanin2023}.

It has also been theoretically shown \cite{Perminov2023,perminov2024multiresonator}
and recently experimentally demonstrated \cite{matanin2025} that an increase in the service life of such QM without a decrease in efficiency can be achieved by using controlled disconnection of this memory from the external waveguide.
Other variants of multiresonator QM are also being developed \cite{solak2025high}, promising to achieve high efficiency.
It is also important to note that the use of QM with a small number of  miniresonators indicates the possibility of implementing spectral-topological QM \cite{perminov2019spectral}.
A key difference between this QM and single-resonator QM is the ability to achieve extremely high efficiency over a wider spectral frequency range  than single-resonator QM by using the specially defined frequencies of these miniresonators and their coupling constants to the common resonator.

The multiresonator QM is becoming an interesting physical system, since in it the quantum state of the signal pulse can be preserved for a long time in a system of spatially remote resonators. 
The study of the quantum properties of such a physical system is made possible by using individual control of the fields stored in each resonator. 
In addition to the fundamental interest in similar quantum states \cite{lange2018entangl}, the availability of individual control of each resonator  opens up new possibilities for using such QM in quantum information processing.
An important step in improving the capabilities of the  multiresonator QM is the development of methods for using atomic ensembles in it. 
This will increase both its lifetime and the number of stored photonic qubits.  
This hybrid QM  will have a number of possibilities in controlling the interactions of photons with atoms, thanks to the use of strong coupling between neighboring miniresonators and with an external waveguide. 
The creation of such QM circuits is now becoming possible thanks to the advent of hybrid integrated quantum photonic circuits \cite{elshaari2020hybrid,bose2024anneal}.

In the first work devoted to the development of such a hybrid QM scheme \cite{perminov2018superefficient}, it was shown that it is possible to achieve high efficiency by using atomic ensembles with fewer atoms compared to the generally accepted impedance matched single-resonator scheme \cite{Moiseev2010cavity,Afzelius-PRA-2010}.
The use of several atomic ensembles in a multiresonator circuit complicates the analytical analysis, therefore, these results were obtained numerically and many general patterns  and conditions for the effective implementation of the storage and retrieval of signal light pulses remained unexplored.
The improvement of integrated technologies in the manufacture of multiresonator optical circuits  \cite{Astratov2017} raises interest in the development of QM based on atomic ensembles in a system of ring resonators combined in a compact circuit  \cite{Haechan_APL2025}. 
In this recent work, the authors studied in more detail the operating modes of such QM using numerical methods and showed its  practical advantages in working with broadband light fields.
However, in the theoretical study \cite{Haechan_APL2025}  of the light field transfer into an atomic system,  the authors neglected the influence of the inhomogeneous broadening of the resonant transition of atoms within each miniresonator. 
In this work, I demonstrate that these properties of the atomic ensemble have an important impact on the performance of this QM.
The significant influence of spectral dispersion  is also shown, which arises due to the use of multiresonator QM, and demonstrated how its compensation can achieve high efficiency.

The present work is devoted to the construction of the theory of QM on atomic ensembles in a multiresonator scheme and obtaining analytical results  which provide detailed information on the  physical processes and mechanism of functioning of such QM, as well as analyzing the influence of its basic parameters on the main properties in storing signal light fields.
The aim of this theory is also to develop a working model of the studied QM, which could be experimentally implemented.
The developed theory analytically explains the results obtained numerically in previous works \cite{perminov2018superefficient,Haechan_APL2025}.
The obtained results also make it possible to determine the influence of the physical parameters of the system of atoms and resonators, at which the studied QM can demonstrate the best operation and maximum efficiency.

\section{Physical model and basic equations }
\label{sec::math_formalism}

Similar to works \cite{perminov2018superefficient,Haechan_APL2025}, we consider a multiresonator QM scheme, including a system of miniresonators, which contain atomic ensembles where all miniresonators are connected to an external waveguide through a common resonator. 
Note that the use of a common resonator facilitates the operation of such a QM, allowing the controlled connection of this resonator to an external waveguide without introducing additional noise into the quantum state of the light, which is stored in a system of miniresonators remote from the switch.
The developed QM is based on the generalization of the principles of spin (photon) echo \cite{Hahn1950,Kurnit1964} into  quantum systems that provide reversible dynamics not only of resonant spin (atomic) ensembles, but also of electromagnetic (light) fields interacting with them, as it was first demonstrated in the QM protocol \cite{moiseev2004photon} for the CRIB protocol and later its implementation was developed in a resonator under the impedance matching condition  \cite{Moiseev2010cavity} and similarly for the AFC protocol \cite{Afzelius-PRA-2010}.

\begin{figure}
\includegraphics[width=0.8\linewidth]{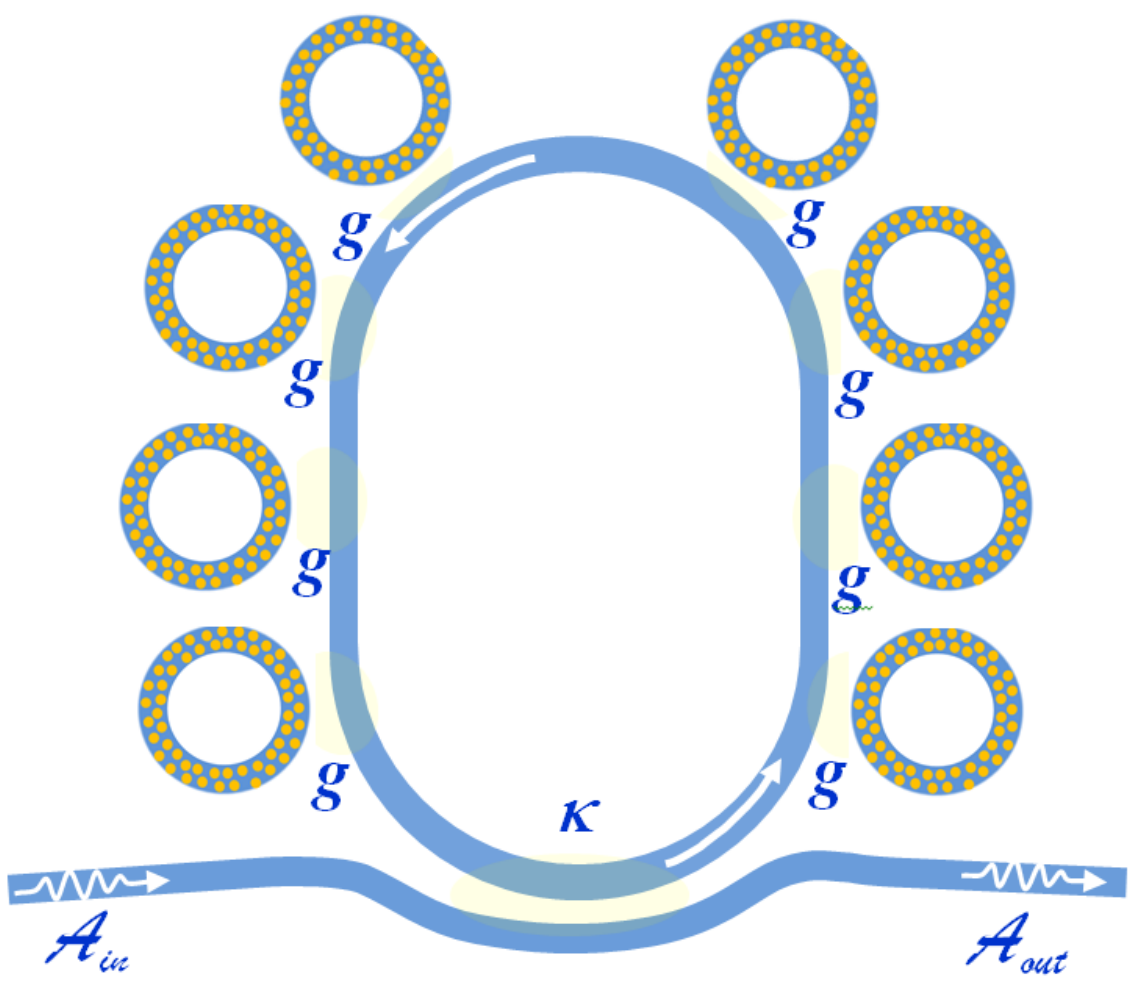}
\caption{\label{QM_scheme} 
Spatial scheme of hybrid multiresonator QM with atomic ensembles: $\kappa$ and $g$ are the coupling constants of  a common (blue) ring resonator with a waveguide and with 8 ring miniresonators; $A_{in}$ and $A_{out}$ are the input and output fields; 
yellow dots in the  miniresonators represent atomic ensembles located in them; $j$-th atom interacts with the mode of $m$-th miniresonator with a coupling constant $f_{j,m}$. 
}
\end{figure}

\begin{figure}
\includegraphics[width=1.0\linewidth]{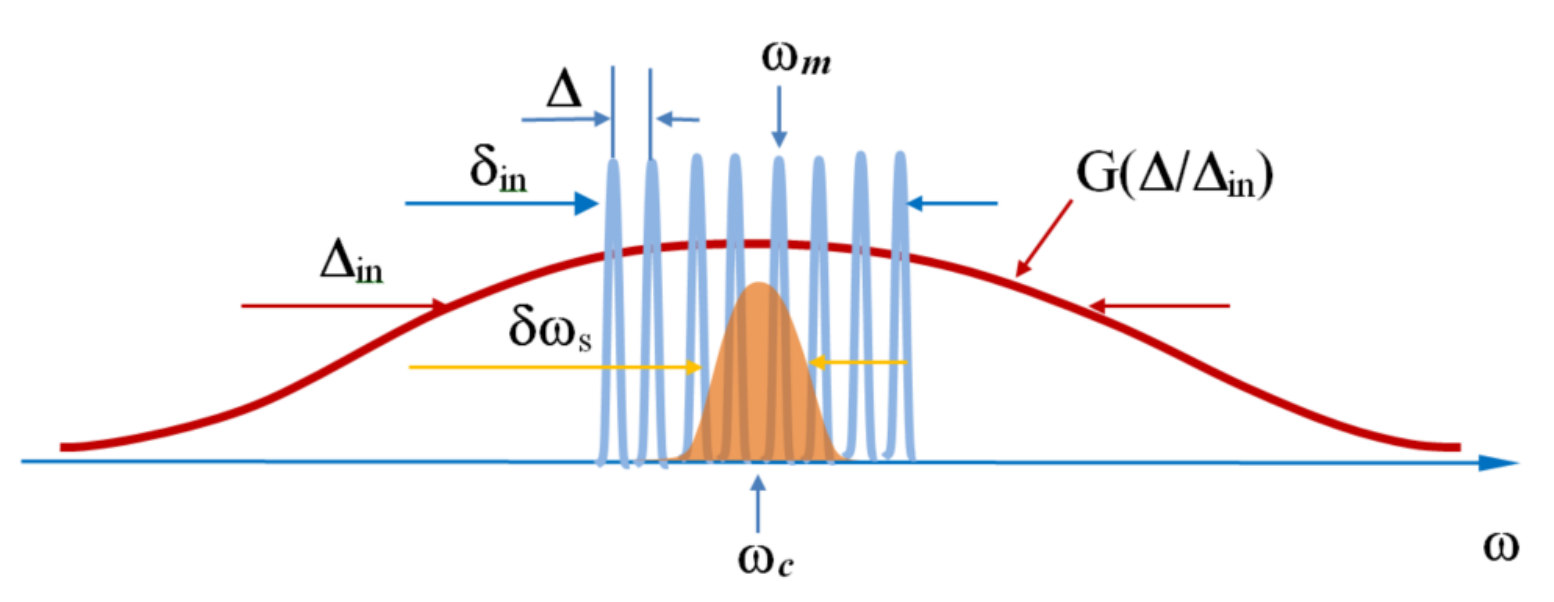}
\caption{\label{Atomic-miniresonator_spectrum} 
Spectrum of inhomogeneously broadened atomic ensemble (line shape $G(\Delta/\Delta_{in})$), miniresonators and signal pulse; $\Delta$ is spectral distance between nearest frequencies of miniresonators, $\Delta_{in}$, $\delta_{in}$ and $\delta\omega_s$  are the spectral widths of atomic ensembles, miniresonators and signal pulse.   
}
\end{figure}

The schematic diagram of the QM under study is shown in Fig.\ref{QM_scheme}. Identical atomic ensembles are placed in several miniresonators.
The frequencies of the miniresonators form a periodic frequency comb 
$\omega_m=\omega_c + \Delta_m$,  where $\Delta_m=m\Delta$ ($m=0,\pm1,\pm2,...$, M is the number of miniresonators)
similar to  AFC protocol of QM \cite{deRiedmatten2008} and realized microwave multiresonator QM \cite{Matanin2023}.
With a large number of miniresonators, we have equality with high accuracy $\delta_{in}\cong M\Delta$. 
However, with a small number of miniresonators $M$, the determination of the spectral width of the QM requires special consideration, since it will be necessary to additionally take into account the spectral neighborhoods near the extreme left and right frequencies. 
In order not to go into the technical details, 
we will keep the original definition $\delta_{in}\equiv M \Delta$.
The central frequency of the optical transition of atoms is tuned to the center of the frequency combs $\omega_c$ of the miniresonators. 
At the same time, we assume that the line of optical transition of atoms is inhomogeneously broadened, the spectral width of which ($\Delta_{in}$) is much wider than the spectral width ($\delta_{in}$) of the miniresonator system ($\Delta_{in}\gg\delta_{in}$).
Thus, the identical ensembles of atoms are placed in all the miniresonators having different frequencies,  located inside an inhomogeneous broadening of the atomic transition line.
Fig. \ref{Atomic-miniresonator_spectrum} shows the arrangement of the miniresonator lines, the inhomogeneously broadened line of atomic transition, and the spectrum of the signal pulse (see also Fig.\ref{Long-lived_storage} below).

The basic Hamiltonian of the system under consideration is written as follows (in units $\hbar$):
\fla{
\hat{H}= & \sum_{m=1}^{M} \Bigl\{(\omega_c+\Delta_m) \hat{b}_m^{\dagger} \hat{b}_m+\frac{1}{2} \sum_{j=1}^{N_m}(\omega_c+\delta_{j,m}) \hat{S}_z^{j,m}\Bigr\} + 
\nonumber 
\\
&\Big( \sum_{m=1}^{M} g_m \hat{b}_m\hat{a}^{\dagger}+ 
\sum_{j=1}^{N_m} f_{j,m}\hat{b}_m\hat{S}_{+}^{j,m}+
H.C.\Big)+
\nonumber
\\
&\omega_c \hat{a}^{\dagger}\hat{a}+\Big(\int d\omega \varrho(\omega)\hat{c}_{\omega}\hat{a}^{\dagger}+ H.C.\Big),
\label{Hamiltonian}
}
where 
$S_{+}^{j,m}=\ket{3_{j,m}}\bra{1_{j,m}}$ and   $S_{-}^{j,m}=\ket{1_{j,m}}\bra{3_{j,m}}$ are 
 raising and lowering atomic operators at the  transition $\ket{1}\leftrightarrow\ket{3}$ of $j$-th atom situated in $m$-th cavity, 
and  $S_{z}^{j,m}=\left(\ket{3_{j,m}}\bra{3_{j,m}}-\ket{1_{j,m}}\bra{1_{j,m}}\right)$ is the operator of the atomic inversion of this atom;
$\hat{a}$($\hat{a}^{\dagger}$),  $\hat{b}_m$($\hat{b}^{\dagger}_m$) and
$\hat{c}(\omega)$ ($\hat{c}^{\dagger}(\omega)$ are boson operators with commutation operators $[\hat{a}(t),\hat{a}^{\dagger}(t)]=1$, $[\hat{b}_m(t),\hat{b}_{m'}^{\dagger}(t)]=\delta_{m,m'}$ and $\hat{c}_{\omega}(t),\hat{c}^{\dagger}_{\omega'}(t)=\delta(\omega-\omega')$, 
$\omega_c$ is a frequency of common resonator and the carrier frequency of a signal pulse, $\Delta_m$ and $\delta_{j,m}$ are the frequency detunings of $m$-th miniresonator  mode and $j$-th atom in this resonator, $g_m$,  $\varrho(\omega)$ and $f_{j,m}$ are the coupling constants of these modes with each other and with $(j,m)$-atom, $N_m$ is a number of atoms in $m$-th miniresonator.

Using Hamiltonian \eqref{Hamiltonian} with input-output approach of quantum optics \cite{Walls,gardiner2015quantum}, we get the following Heisenberg equations for the atomic and field operators:

\fla{
\frac{\partial}{\partial t} \hat{a} & =  - \frac{1}{2}\left(\kappa+\gamma_c\right) \hat{a}
- i \sum_{m}^{M} g_m \hat{b}_{m}+\sqrt{\kappa} \hat{a}_{in}+\hat{F}_c(t),
\label{eqfa-1}
\\
 \frac{\partial \hat{b}_{m} }{\partial t}& = - \left(i\Delta_{m}+\gamma_b\right)  \hat{b}_{m} 
\nonumber
\\
&-i g_m^{*} \hat{a} 
-i\sum_j  f_{j,m} \hat{S}_{-}^{j,m}+\hat{F}_{m}(t),
\label{eqfa-2}
\\
 \frac{\partial \hat{S}_{-}^{j,m} }{\partial t} & = - \left( i\delta_{j,m}+\gamma_a \right)  \hat{S}_{-}^{j,m}  
-i f_{j,m}^{*} \hat{b}_m + \hat{F}_{j,m},
\label{eqfa-3}
}

\noindent
where $\hat{a}_{in}$ is the input signal (with commutation relations $[\hat{a}_{in}(t'),\hat{a}_{in}^{\dagger}(t)]=\delta(t'-t)$) and we also introduced the decay constants $\gamma_c$, $\gamma_b$, $\gamma_a$ and the corresponding  Langivin forces $\hat{F}_{c}(t)$, $\hat{F}_{m}(t)$, $\hat{F}_{j,m}(t)$ by taking into account the additional weak interactions of the resonators and atoms with their environments.
We also take into account that the signal pulse is weak, so the population of the exciting atomic levels is negligible ($\langle\hat{S}_z^{j,m}\rangle\cong -1$).

\subsection{General dynamics and impedance matching conditions}

\textbf{Atoms and resonators at the storage stage}.

Taking into account the linear nature of the Eqs.\eqref{eqfa-1}-\eqref{eqfa-3} 
and focusing primarily on finding the condition for achieving maximum QM efficiency with a weak influence of decoherence, we will limit our consideration to the average values of atomic and field operators
($S_{-}^{j,m}=\langle\hat{S}_{-}^{j,m}\rangle$, $b_{m}=\langle\hat{b}_{m}\rangle$, $a =\langle\hat{a}\rangle$, $A_{in} =\langle\hat{a}_{in}\rangle$) and omit the Langevin forces further.
In the process of solving the system of Eqs. \eqref{eqfa-1}-\eqref{eqfa-3}, we will specify the parameters of the model under study, the constant interactions used in it, the frequencies of the miniresonators, and the nature of the inhomogeneous broadening of the atomic transition. 
To find a solution to the system of equations, we begin by writing down a formal solution for the coherence of $j$-th atom in $m$-th miniresonator $\hat{S}_{j,m} (t)$ and substituting it in Eq.  \eqref{eqfa-2}:

\fla{
S_{-}^{j,m} (t) = &
-i f_{j,m}^{*} \int_{-\infty}^t dt' b_m (t')\cdot
\nonumber
\\ 
& \cdot \exp\{-\left( i\delta_{j,m}+\gamma_a \right) (t-t')\},
\label{atom-coherence}
}
\fla{
& \frac{\partial {b}_{m} }{\partial t} =  - \left(i\Delta_{m}+\gamma_b\right)  \hat{b}_{m} 
-i g_m^{*} \hat{a}
\nonumber
\\
& -  N_m |f_m|^2 
  \int_{-\infty}^t dt' b_m (t')
  \exp\{-\left( \Delta_{in}+\gamma_a \right) (t-t')\},
 \label{eqfa-5}
}

\noindent
where, when calculating the total response of $N_m$ atoms in $m$-th resonator, we switched to a continuous distribution of atomic frequencies  and took into account a large inhomogeneous broadening of the atomic transition $\Delta_{in}$, at which we can take it in the form of a Lorentzian distribution by transferring from discrete ($\delta_{jm}$) to continuous frequency variables ($\omega_{m}$):

\fla{
\sum_j |f_{j,m}|^2 S_{-}^{j,m}(t) \rightarrow
N_m f_m^2 \int d\omega_m  \frac{\Delta_{in}}{\pi(\Delta_{in}^2+\omega_{m}^2)}
S(\omega_{m},t).
\label{sum_int}
}

\noindent
Using the Fourier transform for the mode amplitudes ${b}_{m} {(t)}=\frac{1}{\sqrt{2\pi}}\int d\omega \tilde{b}_{m} {(\omega)}e^{-i\omega t} $ and $a(t)=\frac{1}{\sqrt{2\pi}}\int d\omega \tilde{a} {(\omega)}e^{-i\omega t}$
and taking into account the small spectral width of the signal pulse comparing to the  inhomogeneous broadening $\delta\omega_s\ll\Delta_{in}$, we get the solution of Eq.\eqref{eqfa-5} (see Appendix, Sec.A):

\fla{
 \tilde{b}_{m}{(\omega)}\cong &
-i g_m^{*}\frac { 1}
{\Gamma_{\Sigma,m}+i[\Delta_m-(1-\tilde{\chi})\omega]}\tilde{a} {(\omega)}=
\nonumber \\
&-i g_m^{*} \mathcal{R}_m(\Delta_m,\omega)\tilde{a} {(\omega)},
\label{eqfa-5-dop}
}
where $\tilde{\chi}=\frac{N_m f_m^2 }{\Delta_{in,a}^2}\ll 1$, $\Delta_{in,a}=\Delta_{in}+\gamma_a$, 
$\Gamma_{\Sigma,m}=\Gamma_{a,m}^0+\gamma_b$,
$\Gamma_{a,m}^0=\frac{N_m |f_m|^2}{\Delta_{in,a}}$ is  the relaxation constant, which  is a result of collective interaction of atoms with the resonator mode.
Taking into account that the duration of the signal pulse is significantly shorter than the lifetime of the optical quantum coherence of atoms 
$\delta t_s \ll \gamma_a^{-1}$, we find the following solution \eqref{atom-coherence} for atomic coherence after absorption of the signal pulse:
\fla{
 S_{-}^{j,m} (t)  &\cong
-i \sqrt{2\pi} f_{j,m}^{*} 
e^{ -(i\delta_{j,m}+\gamma_a) t}
\tilde{b}_{m,s}(\delta_{j,m}).
\label{atom-coherence-storage}
}
As can be seen from Eqs. \eqref{atom-coherence-storage}, 
\eqref{eqfa-5-dop}, the spectral dependence of the probability of excitation of j-th atom ($| S_{-}^{j,m} (t)|^2$) in m-th miniresonator is proportional to the Lorentzian line having the linewidth $\Gamma_{\Sigma,m}$ and   decreasing with spectral detuning  from the frequency of m-th miniresonator $|\Delta_m-\omega|$ (where $(1-\tilde{\chi})\omega\cong \omega$.
Thus, in each miniresonator, atoms are excited within a spectral range significantly exceeding the intrinsic spectral width of the miniresonator ($\Gamma_{\Sigma,m}\gg\gamma_b$).
This issue is discussed more precisely below and $| S_{-}^{j,m} (t)|^2$ is shown in the  Fig. \ref{m-th-atomic-excitaions}. 

Substituting the obtained solution \eqref{eqfa-5-dop} into Eq.\eqref{eqfa-1}, we find its solution in the general form: 


\fla{
\tilde{a}(\omega)&=
T_{cw}
(\omega)\tilde{A}_{in}(\omega)= 
\nonumber
\\
&\frac{\sqrt{\kappa}\tilde{A}_{in}(\omega)}
{\left(\frac{1}{2}(\kappa+\gamma_c)
-i\omega 
+
g^2 \int d\Delta_m 
G_r(\frac{\Delta_m}{\delta_{in}})
\mathcal{R}_m(\Delta_m,\omega)
\right) },
\label{a-omega}
}
where $G_r(\frac{\Delta_m}{\delta_{in}})$ is a form factor describing the frequency distribution of miniresonators

\fla{
G_r(\frac{\omega'}{\delta_{in}})=Lim_{\gamma_b\rightarrow 0}\sum_{m=1}^M \frac{|g_m/g|^2 \gamma_b }{\pi(\gamma_b^2+(\omega'-\omega_m)^2)},
\label{G_r_M}
}
 where $\omega_m=\omega_c + \Delta_m$, M is a number of miniresonators, $\delta_{in}\cong M\Delta$, 
 the line width $\gamma_b$ of the miniresonators is already taken into account in the solution \eqref{eqfa-5-dop}, so the limit $Lim_{\gamma_b\rightarrow 0}$ is taken in expression \eqref{G_r_M}. 
 There are no restrictions on the number $M$ of miniresonators in the solution \eqref{a-omega}.
Particular attention should be paid to the study of cases with a small number of miniresonators ($M\sim 3,4,5$) with certain eigen  frequencies and coupling constants ($\sqrt{N}f_m$ and $g_m$) where new interesting properties should be expected due to the appearance of spectral-topological states, the appearance of which in the absence of atomic ensembles \cite{perminov2019spectral} made it possible to achieve superefficient storage of broadband light fields.

The use of general solutions \eqref{eqfa-5-dop} and \eqref{a-omega} in the analytical study of the dynamics of storing and retrieval of signal fields can be significantly simplified by taking into account the spectral properties of signal fields.
Here 
special attention should be also paid to the influence of the spectral parameters of the miniresonators. 
Namely, it is well known that the comb-like nature of this spectrum can manifest itself in the appearance of multiple echo signals, which has been studied repeatedly in AFC protocols of QM (see the reviews \cite{Chaneliere2018,Guo2023,2024-Moiseev-Physics-Uspekhi}).
In this paper, we will primarily focus on the storage stage of short signals in the considered QM scheme. 
During this stage, the narrow lines of the miniresonators do not have time to manifest themselves, moreover, the spectral interval between the nearest frequencies of the miniresonators $\Delta$ also reveals itself only after a time interval significantly exceeding the duration of the signal pulse $\tau=2\frac{\pi}{\Delta}\gg\delta t_s$.
In this case, the dynamics of the interaction of the signal pulse with the QM is influenced by the envelope of the spectrum of the resonance lines of atoms and miniresonators.


The large inhomogeneous broadening of the atomic transition ($\Delta_{in}\gg\delta_{in}$ where $\delta_{in}$ is a linewidth of multi-resonantor system, see Fig. 2) allows using the same atomic ensembles in each  miniresonator.
Assuming the atomic ensembles in each of the resonators to be identical (i.e. $\Gamma_{a,m}^0=\Gamma_{a}^0$),
we will have for the total decay constant of each miniresonator mode $\Gamma_{a,m}^0+\gamma_b\equiv
\Gamma_{a}^0+\gamma_b=\Gamma_{\Sigma}$ ($\Gamma_{a}^0=\frac{N_a f_a^2}{\Delta_{in,a}}$, $N_a$ is a number of atoms in each miniresonator $f_a$ is an average coupling constant).

In carrying out further calculations, we will focus on studying the temporal dynamics at times shorter than the time of automatic rephasing of miniresonators $
\tau=\frac{2\pi}{\Delta}$ . 
In such short time intervals ($t\ll\tau$) of the behavior of the system under consideration, it is possible to switch from a discrete frequency distribution $G_{r} (\Delta_m/\delta_{in})$ of resonators to their continuous distribution $\tilde{G}_{r} (\Delta_m/\delta_{in})$.
(see, for example,  mathematical description of similar transition in  \cite{moiseev2012rephasing}).
Accordingly, bearing in mind the finding of the field of the common resonator $a(t)$ at these time intervals, in the expression \eqref{a-omega}, we replace the sum with the integral over the frequencies of the miniresonators $\tilde{G}_{r} (\frac{\Delta_m}{\delta_{in}})\rightarrow M \tilde{G}_{r} (\frac{\Delta_m}{\delta_{in}})$ (see Appendix A).

There are two interesting options for describing the frequency distribution of miniresonators \eqref{G_r_M}.
In the first case, which takes place in real experiments, we have  miniresonators that form a periodic comb of natural frequencies with a distance $\Delta$ between the nearest  frequencies.
The frequency distribution envelope has a rectangular shape with a total spectrum width
$\delta_{in}$  
(i.e. $\tilde{G}_{r}(\Delta_m/\delta_{in})=\tilde{G}_{r,1}(\Delta_m/\delta_{in})=\frac{1}{\delta_{in}} $ for $|\Delta_m|\leq\frac{\delta_{in}}{2}$ and $\tilde{G}_{r,1}(...)=0$ for $|\Delta_m|>\frac{\delta_{in}}{2}$). 
When the spectral width of the signal spectrum  is less than the spectral width of the miniresonators ($\Delta\ll\delta\omega_{s}\ll\delta_{in}$), the description of the frequencies of the miniresonators can be also simplified by using the Lorentzian form 
($\tilde{G}_{r}(\Delta_m/\delta_{in})=\tilde{G}_{r,2}(\Delta_m/\delta_{in})=\frac{\delta_{in}}{\pi(\delta_{in}^2+\Delta_{m}^2)}$), which simplifies the mathematical description, leading to almost the same analytical result. 
In a further analysis, we will compare the results of these two descriptions.

Taking into account the continuous distribution of the frequencies of the miniresonators in the Eq. \eqref{a-omega}, after carrying out calculations we obtain

\fla{
\tilde{a}(\omega)=&
T_{cw}
(\omega)\tilde{A}_{in}(\omega)= 
\nonumber
\\
&\frac{\sqrt{\kappa}\tilde{A}_{in}(\omega)}{\left(\frac{1}{2}(\kappa+\gamma_c)+\frac{M g^2 }{{\delta_{in} \mathcal{F}_{1,2}(\chi\delta_{in},\chi \Gamma_{\Sigma},\omega)}} -i\omega \right) },
\label{a-omega-first}
}
where $\mathcal{F}_{1}(...)$ and $\mathcal{F}_{2}(...)$ correspond to two variants of miniresonator frequency distribution:

\fla{
\mathcal{F}_{1}(...)\cong &
\frac{1}{\left(\pi-2\arctan\left(\frac{2\Gamma_{\Sigma}}{\delta_{in}} \right)+i \frac{4 \delta_{in}\frac{\omega}{\chi}}{ (\delta_{in}^2+4\Gamma_{\Sigma}^2+4(\frac{\omega}{\chi})^2}\right)},
\label{F-1}
\\
\mathcal{F}_{2}(...)= &\left(
1+\frac{\chi\Gamma_{\Sigma} -i\omega}{\chi\delta_{in}}
\right).
\label{F-2}
}
The exact solution for the function $\mathcal{F}_{1}(...)$ is presented in the Appendix A. 
The solution \eqref{F-1} takes place 
in the limit of narrow band signal pulse $\delta\omega_s\ll\delta_{in}$. 
The appendix also shows that the  which is completely analogous to the function $\mathcal{F}_{2}$ with Lorentzian shape in Eq. \eqref{F-2} for arbitrary spectral width of a signal pulse if ${\Gamma_{\Sigma}\ll\delta_{in}}$ (where $\frac{4}{\pi}\Gamma_{\Sigma}$ should be replaced only  by $\Gamma_{\Sigma}$).

The Eq. \eqref{a-omega-first} is the first main result obtained, which only takes into account the assumption of a large inhomogeneous broadening of the atomic transition ($\Delta_{in}\gg\delta t_s^{-1},\delta_{in}$).
Below, using the solution \eqref{a-omega-first} we  analyze the QM dynamics and determine the physical parameters providing  high efficiency and broadband storage  of a signal light pulse.
Applying the relation connecting the input and output fields with the field of a common resonator \cite{Walls}  $A_{in}(t)+A_{out}(t)=\sqrt{\kappa} a(t)$, we get 

\fla{
\tilde{A}_{out} (\omega)=&U(\omega) \tilde{A}_{in}(\omega)=
\nonumber
\\
=& \frac{\left(\frac{1}{2}(\kappa-\gamma_c)-\frac{M g^2 }{\delta_{in} \mathcal{F}_{1,2}(\chi\delta_{in}, \chi\Gamma_{\Sigma},\omega)}+i\omega \right)}{\left(\frac{1}{2}(\kappa+\gamma_c)+\frac{M g^2 }{\delta_{in} \mathcal{F}_{1,2}(\chi\delta_{in},\chi \Gamma_{\Sigma},\omega)} -i\omega \right) } \tilde{A}_{in}(\omega).
\label{a-out}
}

\textbf{Basic impedance matching condition}.
\\
By taking the carrier frequency of the signal pulse equal to zero $\omega=0$ in the Eq. \eqref{a-out}, we obtain the impedance matching condition:  

\fla{
\kappa=\gamma_c+\frac{2M g^2 }{\delta_{in} \mathcal{F}_{1,2}(\chi\delta_{in},\chi \Gamma_{\Sigma},0)},
\label{impedance-m}
}
which provides  close to 100\%  transfer of an input signal to the system of resonators $+$ atoms. 
Comparing  \eqref{impedance-m}  with the solution for multiresonator QM without using atomic ensembles \cite{Matanin2023}, we find that the interaction of miniresonators with atomic ensembles is manifested in  \eqref{impedance-m} by  increasing  the total spectral width of the frequency comb of miniresonators ($\delta_{in}\rightarrow \delta_{in} \mathcal{F}_{1,2}(\chi\delta_{in},\chi \Gamma_{\Sigma},0)>\delta_{in}$) (where $\mathcal{F}_{1,2}(\chi\delta_{in}, \Gamma_{\Sigma},0)>1$ and it is  a growing function of $\Gamma_{\Sigma}$ see Eqs.\eqref{F-1}, \eqref{F-2}).
Thus, the interaction with the atomic ensembles  reduces the coupling constant $\kappa$ required to satisfy the impedance matching condition. This effect increases with  ratio $\frac{\Gamma_{\Sigma}}{\delta_{in}}$.

Let's take a closer look at it for the first  (rectangular) variant of the miniresonator frequencies. Here we get from Eq. \eqref{impedance-m}:

\fla{
\kappa &=\gamma_c+
2\pi \frac{g^2}{\Delta}
\left(1-\frac{2}{\pi}\arctan\left(\frac{2\Gamma_{\Sigma}}{\delta_{in}} \right)\right).
\label{impedance_match}
}

This solution is valid for arbitrary parameters included in it and shows that enhancing the interaction of miniresonators with atomic ensembles reduces the coupling constant of the common resonator with the waveguides $\kappa$, which is necessary for impedance matching condition.
The following three limiting cases are interesting: 

\textbf{1)} 
The case of weak interaction between  atomic ensemble with miniresonator ($\frac{\Gamma_{\Sigma}}{\chi\delta_{in}}<1 $):

\fla{
\kappa \cong \gamma_c+
2\pi \frac{g^2}{ \Delta}
\left(1- \frac{4\Gamma_{\Sigma}}{\pi\delta_{in}} \right),
\label{impedance_match-small_Gamma}
}
which transfers to impedance matching condition of empty miniresonators \cite{Matanin2023}  for $\frac{4\Gamma_{\Sigma}}{\pi\delta_{in}}\ll 1$.  

\textbf{2)} Strong interaction of atoms with miniresonator ($\frac{2\Gamma_{\Sigma}}{\delta_{in}}\gg 1 \rightarrow $ $\arctan[\frac{2\Gamma_{\Sigma}}{\delta_{in}}] \approx \frac{\pi}{2}-\frac{\delta_{in}}{2\Gamma_{\Sigma}}$): 

\fla{
\kappa \cong \gamma_c+
2\pi \frac{g^2}{ \Delta}
\frac{\delta_{in}}{\pi \Gamma_{\Sigma}}=
\gamma_c+ \frac{2M g^2}{\Gamma_{\Sigma}},
\label{impedance_match-large_Gamma}
}
where $\frac{2M g^2}{\Gamma_{\Sigma}}\ll \frac{2\pi g^2}{\Delta}$, which describes the interaction of a common resonator with identical $M$ mini  resonators having a spectral width $\Gamma_{\Sigma}$ an increase in which leads to a narrowing of the working spectral range.
For the effective implementation of broadband QM, it is necessary to limit the increase of $\Gamma_{\Sigma}$ in order to preserve the condition $\frac{M g^2}{\Gamma_{\Sigma}} \gg \gamma_c$.

\textbf{3)} miniresonators with Lorentzian frequency distribution. 
Using Eqs. \eqref{F-2} and \eqref{impedance-m}, 
 we get the following impedance matching condition:   

\fla{
\kappa  =  \gamma_c+
\frac{2M g^2 }{ \left(
\delta_{in}+\Gamma_{\Sigma}\right)}.
\label{impedance_match-small_Gamma}
}

Thus, the well-known impedance matching  condition \eqref{impedance_match-large_Gamma}  is obtained only if $ \delta_{in}\gg \Gamma_{\Sigma}$.
At the same time,  when the signal's spectral width is sufficiently small ($\delta\omega_s\ll\delta_{in}$), using the Lorentzian shape yields exactly the same result. Therefore, when studying the properties of optical quantum noise in the studied QM scheme, we will use the Lorentzian line shape, which significantly simplifies the analytical calculations.

Below we analyze in detail the conditions for implementing highly efficient storage of the signal pulse, and the influence of the interaction of miniresonators with atoms on this process.

\subsection{Spectral impedance matching condition}

\begin{figure}
\includegraphics[width=0.8\linewidth]{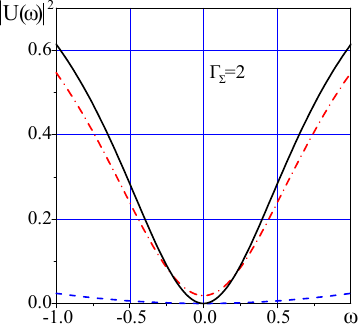}
\caption{\label{Output_field-1} 
Behavior of the spectral reflection function $|U(\omega)|^2$ depending on  $\omega=2\pi\nu$ ($-1\leq \omega \leq 1$): the red dashed-dotted line corresponds to the reflection under the impedance matching condition \eqref{impedance_match}, neglecting in this condition the interaction of miniresonators with atoms;
black solid line - taking into account in the condition \eqref{impedance_match}  the interaction of miniresonators with atomic ensembles.  where $\kappa=1$, $\delta_{in}=10$, $\Gamma_{\Sigma}=2$.
For the convenience of demonstrating the spectral behavior of all the graphs, the same frequency unit  in all figures, which is taken for calculating the graph with red dashed-dotted (and black solid) line. 
All the graph  are presented in the same units, but for different parameters of the system under study (the same relation are used in other figures);
blue dashed line - taking into account two impedance matching conditions \eqref{impedance_match} and  \eqref{broaderning_condition} where $\Gamma_{\Sigma}=2$, $\delta_{in}=10$, $\kappa=15.51$. 
}
\end{figure}

Returning again to the solution that writes the spectrum of reflected radiation \eqref{a-out}, we get spectral impedance matching condition:

\fla{
\frac{4M g^2}{\chi(\delta_{in}^2+4\Gamma_{\Sigma}^2)}=1.
\label{broaderning_condition}
}
Eq.\eqref{broaderning_condition} is obtained by substituting the Eq.\eqref{F-1} into Eq. \eqref{a-omega} (and in Eq. \eqref{a-out}) and taking into account that $\delta_{in}\gg\omega$, we equate to zero  the term proportional to the first degree in $\omega$, which leads to Eq.\eqref{broaderning_condition}.
If it is satisfied, the transfer function $U(\omega)$ (and $T(\omega)$ in Eqs.\eqref{a-omega}) will only depend quadratically on the frequency offset $\omega$ leading to the 
 broadening of the working spectral width of the QM  similar to the QM for atoms in a single resonator \cite{Moiseev2010cavity}.
 Taking spectral matching into account in Eq. \eqref{impedance-m} leads to the following expression for the basic impedance matching condition:

\fla{
\frac{\kappa}{2} =
&\frac{\gamma_c}{2}+\chi\frac{\delta_{in}^2+4\Gamma_{\Sigma}^2}{4\delta_{in}}\left(\pi- 2\arctan\left(\frac{2\Gamma_{\Sigma}}{\delta_{in}} \right)\right)_{\mid\Gamma_{\Sigma}\ll\delta_{in}}
\nonumber \\
\cong & \frac{\gamma_c}{2}+\frac{\pi}{4}\chi\delta_{in}\left(1- \frac{4\Gamma_{\Sigma}}{\pi\delta_{in}} \right),
\label{impedance-m-3}
}

\noindent
where the case of the predominance of inhomogeneous broadening over the constant of interaction of an ensemble of atoms with miniresonators is also noted (and $\delta_{in}\gg \frac{4}{\pi}\Gamma_{\Sigma}$ is needed).
The graphs in Fig. \ref{Output_field-1} show the behavior of the spectral reflection function $|U(\omega)|^2$ (see Eq. \eqref{a-out}), 
The frequency range where reflection tends to zero corresponds to high efficiency of signal pulse transfer into the atomic ensembles.
Comparison of the graphs (black solid and red dashed-dotted lines) in Fig. \ref{Output_field-1} demonstrates the influence of the interaction of atoms with miniresonators on the impedance matching conditions (see spectral range near $\omega=0$).
The blue dashed line in Fig. \ref{Output_field-1} shows 
that taking into account the spectral matching condition \eqref{broaderning_condition} greatly expands the spectral range of QM compared to the other two graphs.

\begin{figure*}[t]
\includegraphics[width=0.8\linewidth]{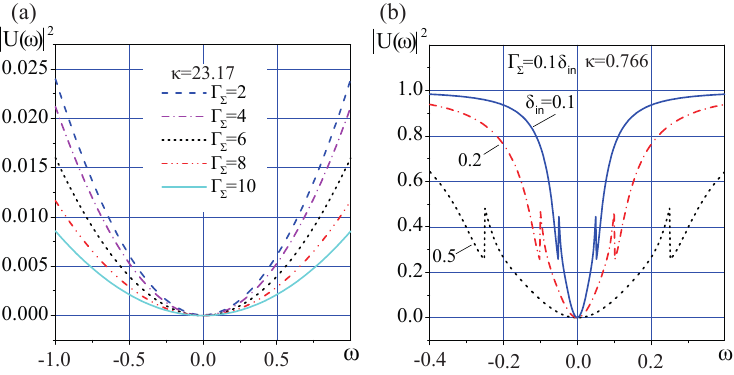}
\caption{\label{Output_field-4-5} 
Spectral reflection function $|U(\omega)|^2$ 
with two impedance matching conditions \eqref{impedance_match} and \eqref{broaderning_condition}:
\textbf{(a)} linewidth $\delta_{in}=10$, 
$\Gamma_{\Sigma}=2,4,6,8,10$ ($\kappa=23.17$).
\textbf{(b)}    $\Gamma_{\Sigma}=0.1\delta_{in}$, linewidth $\delta_{in}=0.1,0.2,0.5$ 
( $\kappa=0.766$). 
}
\end{figure*}

\begin{figure*}
\includegraphics[width=0.8\linewidth]{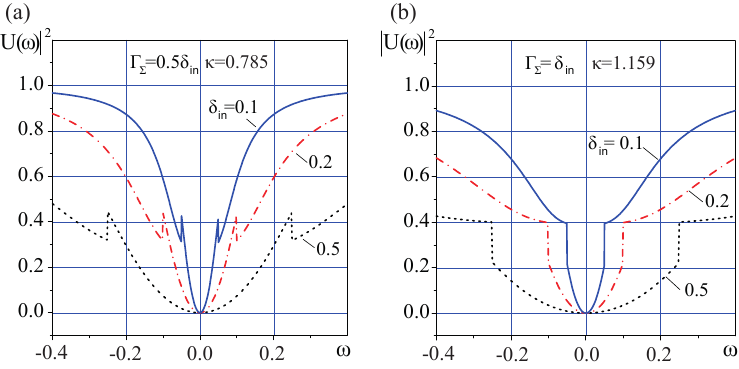}
\caption{\label{Output_field-6-7} 
Spectral reflection function $|U(\omega)|^2$ 
with two impedance matching conditions \eqref{impedance_match} and \eqref{broaderning_condition}:
\textbf{(a)}
 $\Gamma_{\Sigma}=0.5\delta_{in}$, linewidth $\delta_{in}=0.1,0.2,0.5$ 
($\kappa=0.785$),
\textbf{(b)}
 $\Gamma_{\Sigma}=\delta_{in}$, linewidth $\delta_{in}=0.1,0.2,0.5$  ($\kappa=1.159$); for  large $\Gamma_{\Sigma}$ see Eqs.\eqref{F-1-AP} in Appendix A. 
}
\end{figure*}

Fig. \ref{Output_field-4-5}  (a) 
shows graphs of the spectral behavior of reflection for for different strengths of interaction of miniresonators with atomic ensembles.
The graphs  show that an increasing the interaction constant $\Gamma_{\Sigma}$  of miniresonators with atoms can lead to a spectral broadening of the effective light storage (when the interaction constant $\Gamma_{\Sigma}$ increases from $0.2\delta_{in}$ to $\delta_{in}$.
At the same time, the storage  efficiency reaches 99 \%  within the spectral range 0.2 $\delta_{in}$ (see curves with $\Gamma_{\Sigma}$=8 and 10, where  $\kappa=23.17$ with $\delta_{in}=10$).

Fig.  \ref{Output_field-4-5}  (b) and Figs. \ref{Output_field-6-7} (a), (b)
(presented in the same units as in Fig.\ref{Output_field-4-5}  (a),
show the behavior of the spectral reflection function $|U(\omega)|^2$ for different interaction constants $\Gamma_{\Sigma}=0.1\delta_{in},0.5\delta_{in},1.0\delta_{in}$ and spectral widths $\delta_{in}=0.1,0.2,0.5$ of the miniresonators under impedance matching conditions \eqref{impedance_match} and \eqref{broaderning_condition}.  
Fig. \ref{Output_field-4-5}  (b) and Fig.
\ref{Output_field-6-7} (a) show 
how the reflection changes (decreases) dramatically at the spectrum boundary of the miniresonator system.
Note that within the entire spectral range there are resonant atoms. 
Of interest is the presence of some weakening of the reflection at a small detuning beyond the spectrum boundaries of the miniresonators, which is apparently explained by the manifestation of strong frequency dispersion. 
The general tendency to weaken the reflection with an increase in the frequency detuning from the center of the line indicates a weakening of the interaction of the signal pulse with atoms whose spectrum is beyond the spectrum of the miniresonators and a departure from the impedance matching condition.

As can be seen in Fig.
\ref{Output_field-6-7}  (b),
an increase in interaction with atoms $\Gamma_{\Sigma}$ leads to a weakening of the reflection of the signal within the spectrum of miniresonators, which is manifested in a deepening of the minimum of the function $|U(\omega)|^2$.
It is also noteworthy that, in contrast to the weak interaction of atoms with miniresonators, there is no longer a jump in the decrease in signal reflection when going beyond the spectrum of miniresonators.
The graphs presented in the Figs. 
\ref{Output_field-4-5}  (a),(b) and Figs. \ref{Output_field-6-7} (a),(b)
allow us to conclude that the strengthening of the interaction of miniresonators with atomic ensembles leads to an improvement in the spectral properties of QM.
It is also seen that the rectangular nature of the frequency distribution of miniresonators ${G}_{r,1}(\Delta_m/\delta_{in})$ leads to a quadratic dependence on the behavior of the spectral efficiency. 
In this connection, an even greater increase in the working spectral width of QM requires the introduction of a frequency dependence of the interaction constant ($g_m$) of miniresonators with a common resonator, which be a subject of further studies.

Until now we have been interested in the joint manifestations of miniresonators with atoms in the absorption of a signal pulse by a QM  cell.
However, the implementation of long-lived QM involves timely and highly efficient transfer of a signal pulse  to atomic ensembles from miniresonators.
To clarify this process 
it is necessary to consider its spectral and temporal behavior in more detail, taking into account that the system of miniresonators is phased over time $\tau=2\pi/\Delta\gg\delta t_s\sim \delta\omega_s^{-1}$, which can change its interaction with the common resonator.

\subsection{Dynamics of  miniresonators and atoms}

Important information about the properties of the signal pulse absorption can be obtained by sequentially examining the behavior of atomic coherence and field amplitude in $m$-th miniresonator.
By using  Eq.\eqref{a-omega} and \eqref{eqfa-5-dop} we get the spectral response $\tilde{b}_m(\omega)$ of the $m$-th resonator mode to the signal radiation:

\fla{
\tilde{b}_m(\omega) & = - i g_m^{*} \frac{\chi T_{cw}(\omega)}{\left ( \chi\Gamma_{\Sigma}+i(\chi\Delta_{m}-\omega)\right)} 
\tilde{A}_{in}(\omega).
\label{b-m-omega}
}

Taking into account the flat vertex in the spectral dependence of the $T(\omega)$ function, Eq. \eqref{b-m-omega} describes the Lorentzian  shapes  of excitation of the $m$-th mode  with a slightly shifted eigen frequency ($\Delta_{m}\rightarrow\chi\Delta_{m}$) and considerably increased spectral width ($\gamma_b\rightarrow \gamma_b+\Gamma_{a}^0\gg \gamma_b$) due to the interaction with $m$-th atomic ensemble.
Going to the time domain, by taking into account a sufficiently narrow spectral width ($\delta\omega_s\ll\delta_{in}+\Gamma_{\Sigma}$)  according to Eqs.\eqref{a-omega}, \eqref{F-1} and  the impedance matching condition, we get: 
\fla{
{a}(t)\cong\frac{{A}_{in}(t)}{\sqrt{\kappa}}.
\label{a-at-impedance}
}
The solution \eqref{a-at-impedance}  describes the fast transfer of a signal pulse into the system of miniresonators from the common resonator during the temporal duration of a signal pulse ($\delta t_s\sim\delta\omega_s^{-1}$).
Having the solutions Eqs. \eqref{eqfa-5-dop} and \eqref{a-at-impedance}, 
we obtain for the time evolution of the  $m$-th field mode in this case:
\fla{
 b_{m} ( t>\delta t_s) \approx &- i 
 \sqrt{2\pi}\frac{g_m^{*}\chi}{\sqrt{\kappa}}\tilde{A}_{in}(\chi\Delta_{m})
\cdot
\nonumber
\\
&\exp\{-\left( i\chi\Delta_{m}+\chi \gamma_b + \chi\Gamma_{a}^0 \right) t\},
 \label{b-m-time-behavior}
}
which  shows that $m$-th miniresonator captures the spectral component $\chi\Delta_{m}$ of the signal field practically during the duration of the signal pulse $\delta t_s$. 
However, in each of the miniresonators, the field exists for a period of time $\sim (\chi \gamma_b + \chi\Gamma_{a}^0 )^{-1}$, interacting with the atomic system. 
It is important that during this time the dynamics of the atomic system is not negatively affected by decoherence effects in each miniresonator and in atomic ensembles,  that is, the conditions  $\gamma_b/\Gamma_{a}^0 \ll1$ and $\gamma_a/\Gamma_{a}^0 \ll1$ must be hold.
Taking into account that $\Gamma_{a}^0\sim\Delta$  and $\Delta\ll\delta\omega_s\ll\delta_{in}$ ($\Delta$ is a spectral interval between nearest frequencies of miniresonators (see \eqref{atomic_coupling} below), we get $\Gamma_{a}^0\delta t_s\ll 1$.

We can also estimate the behavior of the degree of excitation $P_M(t)$ of the miniresonator system assuming a sufficiently large spectral width of the miniresonator $\delta_{in}\gg \delta\omega_s$.
In this case, without affecting the result of the calculations, we used the Lorentzian distribution of miniresonator frequencies  ${G}_r(\Delta_m/\delta_{in})={G}_{r,L}(\Delta_m/\delta_{in})=\frac{\delta_{in}}{\pi(\delta_{in}^2+\Delta_{m}^2)}$.
Using \eqref{b-m-time-behavior} we get:

\fla{
P_M(t)&= \sum_m^M \abs{b_m(t)}^2
\nonumber \\
&\approx 2\pi \frac{M g^2 \chi^2}{\kappa} 
\exp\{-2\chi\left(  \gamma_b + \Gamma_{a}^0 \right) t\}\cdot
\nonumber \\
& \int \Delta_m \frac{\delta_{in}}{\pi(\delta_{in}^2+\Delta_m^2)} \abs{\tilde{A}_{in}(\chi\Delta_{m})}^2_{\delta_{in}\gg\delta\omega_s}
\nonumber \\
& \approx 2 \frac{M g^2 \chi}{\kappa\delta_{in}} 
\exp\{-2\chi\left(  \gamma_b + \Gamma_{a}^0 \right) t\} W_{in},
\label{all_bm}
}
where $W_{in}=\int_{-\infty}^{\infty} dt|A_{in}(t)|^2$ is a number of photons in an input light pulse. 
Considering also $\delta_{in}
\gg \Gamma_{a}^0$, $\chi\approx 1$ with impedance matching condition  \eqref{impedance_match-small_Gamma} in Eq. \eqref{all_bm}, we get exponential energy loss from the miniresonator system:

\fla{
P_M(t)\approx W_{in}\exp\{-2\chi\left(  \gamma_b + \Gamma_{a}^0 \right) t\}.
\label{miniresonator-energy-1} 
}

The solution \eqref{miniresonator-energy-1} describes the limit case of a very short signal pulse and a very large spectral width of the spectrum of the miniresonators ($\delta_{in}\gg \Gamma_{a}^0$).
The solution \eqref{miniresonator-energy-1},  based on the assumption of very fast signal transmission from the common resonator to the miniresonator system allows us to understand the rate of signal pulse energy transmission from the miniresonator system to the atomic ensembles.
However, it does not take into account the influence  of interaction with atoms on the transfer of the signal pulse to the miniresonator system from the common resonator.
We can obtain more detailed information about the dynamics of the behavior of the miniresonator system using the Eq.  \eqref{eqfa-5-dop}. 
Carrying out similar calculations under the assumption that $\delta_{in}\gg\delta\omega_s$ and $\delta_{in}-\Gamma_{\Sigma}\gg\delta\omega_s$  ($\delta_{in}>\Gamma_{\Sigma}$):

\fla{
P_M(t)&= \sum_m^M \abs{b_m(t)}^2 \cong\frac{2 Mg^2}{\kappa} 
\frac{\delta_{in}}{(\delta_{in}^2-\Gamma_{\Sigma}^2)}\cdot
\nonumber \\
&\{\chi \int_{-\infty}^t dt'\abs{A_{in}(t')}^2 e^{-2\chi\Gamma_{\Sigma} (t-t')} 
-\frac{\abs{A_{in}(t)}^2}{2\delta_{in}}\}.
\label{miniresonator-energy-2}
}

In the limit $\delta_{in}\gg\Gamma_{\Sigma}$ and $\delta\omega_s\gg \Gamma_{\Sigma}$, the solution \eqref{miniresonator-energy-2} comes down to the solution \eqref{miniresonator-energy-1}.
As can be seen from the solution  \eqref{miniresonator-energy-2}, when  deviating from these conditions, the excitation dynamics of the miniresonator changes noticeably at the stage of interaction with the input signal pulse.
After the end of the impulse, the second term in Eq. \eqref{miniresonator-energy-2} disappears and the solution takes the form:

\fla{
P_M(t)& \cong\frac{2 Mg^2 \chi}{\kappa} 
\frac{\delta_{in}}{(\delta_{in}^2-\Gamma_{\Sigma}^2)} e^{-2\chi\Gamma_{\Sigma}t}\cdot
\nonumber \\
& \int_{-\infty}^{t>\delta t_s} dt'\abs{A_{in}(t')}^2 e^{2\chi\Gamma_{\Sigma}t'}.
\label{miniresonator-energy-3}
}

Thus after the signal pulse has entered the miniresonators and its duration is shorter than $\Gamma_{\Sigma}^{-1}$, we see exponential decay of energy captured in all the miniresonators:
All the trapped signal energy is transferred into the atomic ensemble with a rate $ \chi\Gamma_{\Sigma}$ provided that $\Gamma_{a}^0\gg\gamma_b$, which is necessary to minimize irreversible signal pulse losses in miniresonators.
The main part of the light energy will be transferred with high efficiency to atoms during time interval $T$ satisfying the condition $ \Gamma_{a}^0 T \geq 3 $. 
At the same time, if $5$ miniresonators are used, no more than 1.2 $\%$ energy remains in them at time $t=T$.
However, this process can continue as long as the miniresonators are out of phase with respect to each other, i.e. if $T < \tau = \frac{2\pi}{\Delta}$,
for example at $T = \tau/2> 3 (\Gamma_{a}^0)^{-1}$.
Accordingly, we obtain the following stronger condition   

\fla{
\Gamma_{a}^0=\frac{N_a f_a^2}{\Delta_{in,a}}\geq 6/\tau =\frac{3\Delta}{\pi},
 \label{atomic_coupling}
}
providing highly efficient transfer of signal light to atomic ensembles if  $N_a \geq\frac{3  \Delta }{\pi }\frac{\Delta_{in,a}}{f_a^2}$.
At the same time, it is necessary to note here the conditions for the atomic coherence time $T_2=\gamma_a^{-1}$ and the eigen linewidth $\gamma_b$ of the miniresonators: $T_2\gg 2\pi /\Delta$ and $2\pi \gamma_b/\Delta\ll 1$. 
The latter condition, together with the condition for a coupling constant  between resonators and atomic ensembles $\Gamma_{a}^0$, imposes basic requirements both for a high Q-factor of miniresonators ($Q\gg\frac{\pi\omega_c}{\Delta}$, where $\omega_c$ is a frequency of miniresonator). 
Having the relations Eqs.\eqref{impedance-m}, \eqref{impedance_match} and Eq. \eqref{atomic_coupling}, we can estimate how much easier the requirements are for using atomic ensembles for multiresonator QM in relation to single-resonator QM.
The implementation of QM on an atomic ensemble in one resonator implies the fulfillment of the matching condition
$\frac{N_s f_s^2}{\Delta_{in,a}}=\kappa/2$ \cite{Moiseev2010cavity,Afzelius-PRA-2010} ($f_s$ is a coupling constant of atoms with a single resonator) and comparing it with the condition Eq.\eqref{atomic_coupling}.
Comparing the number of atoms $N_a$ with the number of atoms $N_s$ in a QM with a single resonator, we find:

\fla{
N_a\approx
\frac{6\Delta}{\pi\kappa}
\abs{\frac{f_s}{f_a}}^2 N_s,
}
which is at least in $\frac{\pi\kappa}{6\Delta}\gg 1$ times less than $N_s$.
This quantitative estimate, which specifies the observation made earlier in  \cite{perminov2018superefficient}. 
Interestingly, if we assume that the total number of miniresonators $M$ fills a spectral range equal to a $\kappa$ ($M\Delta \cong \kappa$), then the total number of atoms in all resonators will equal the number of atoms in a single-resonator quantum memory.
But in miniresonators, the coupling constant with atoms $f_a$ can be significantly stronger then the constant $f_s$ in a single larger resonator, which will lead to a significant decrease in the total number of atoms $N_a$.

It is also important to keep in mind that increasing the coupling constant of the interaction of atomic ensembles with the miniresonator $\Gamma_{a}^0=\frac{N_a f_a^2}{\Delta_{in}}$ can lead to too large a narrowing of the working spectral range (see Eqs. \eqref{impedance-m}-\eqref{impedance-m-3}).
The implementation of the impedance matching condition is significantly simplified by the fact that it depends to a greater extent on the parameters of the interaction of the miniresonators with the main resonator.
It becomes easy to fulfill this condition when the coupling constants between the resonators $g_m\equiv g$ can be chosen to be large enough.

The next important issue is to clarify the parameters of the exited atomic state resulting from the storage of a signal pulse.
Using Eqs.\eqref{atom-coherence} and \eqref{b-m-omega}, we get for $t \geq T$:

\fla{
 S_{-}^{j,m} (t)  &=
-i \sqrt{2\pi} f_{j,m}^{*} 
e^{- (i\delta_{j,m}+\frac{1}{2T_1})t+i\delta\phi_{j,m}(t)}
\tilde{b}_m(\delta_{j,m}),
\label{atom-coherence-2}
\\
\tilde{b}_m(\omega) & = - i g_m^{*} \frac{\chi T_{cw}(\omega)}{\left ( \chi\Gamma_{\Sigma}+i(\chi\Delta_{m}-\omega)\right)} 
\tilde{A}_{in}(\omega),
\label{b-m-omega}
}
where we have concretized the nature of the decay of atomic coherence, highlighting the appearance of a random phase shift $\delta\phi_{j,m}(t)$  and the manifestation of the transition of the atom to other energy states determined by the time $T_1$ of longitudinal (energy) relaxation ($T_{cw}(\omega)$ see in Eq.\eqref{a-omega}).
Probability of $(j,m)$-th atom to be in excited state will be:

\fla{
|S_{-}^{j,m} (t)|^2  
= e^{- \frac{t}{T_1}}&
|T_{a,m}(\chi\Delta_m-\delta_{j,m})|^2\cdot
\nonumber \\
& |T_{c,w}(\delta_{j,m})|^2
|\tilde{A}_{in}(\delta_{j,m})|^2,
\label{b-m-omega=2}
}
where

\fla{
|T_{a,m}(\chi\Delta_m-\delta_{j,m})|^2=\frac{2\pi \chi^2\abs{fg}^2 }
{\left (\chi \Gamma_{\Sigma} \right)^2+(\chi\Delta_{m}-\delta_{j,m})^2}.
\label{atomic_exit_m_resonator}
}
In the calculations above, we assumed  $f_{j,m}=f_a$, $g_m=g$ (see also Eq.\eqref{a-omega}). 

\begin{figure}
\includegraphics[width=1.0\linewidth]{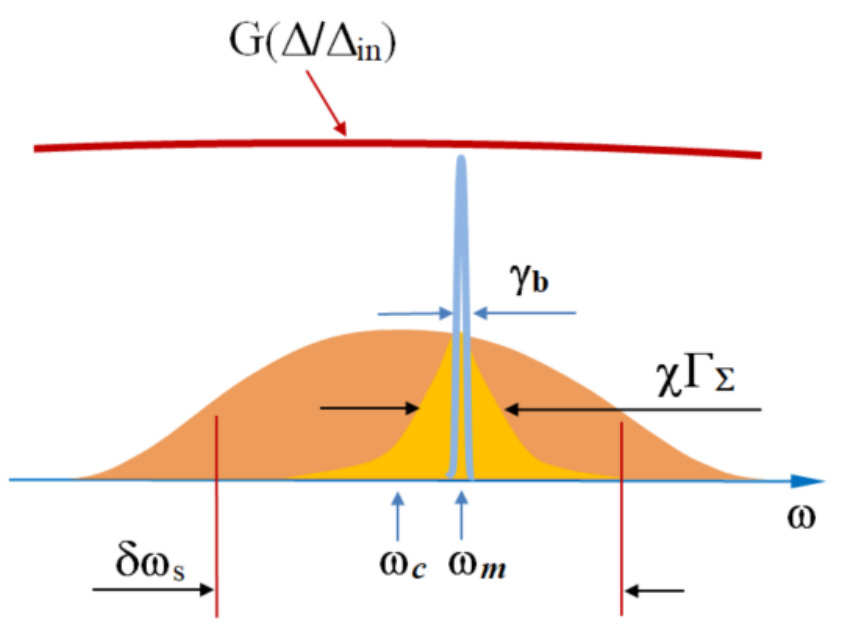}
\caption{\label{m-th-atomic-excitaions} 
Spectrum of excited atoms in $m$-th miniresonator which has Lorentzian shape   with spectral width $\chi\Gamma_{\Sigma}\geq \frac{3\Delta}{\pi}$, $\gamma_b$ and $\delta\omega_s$ are the spectral widths  of the eigenmode of the miniresonator and the signal pulse.
}
\end{figure}

In the physical sense, the transfer function $ |T_{c,w}(\delta_{j,m})|^2$  describes the probability of transferring the frequency mode $\omega=\delta_{j,m}$ of a signal photon from the waveguide to the common resonator. 
When both impedance matching conditions of agreement are met $ |T_{c,w}(\delta_{j,m})|^2\cong \frac{1}{\kappa}$.
The function $|T_{a,m}(\chi\Delta_m-\delta_{j,m})|^2$ describes the probability of transferring this spectral mode from the common resonator to the $j$-th atom of the $m$-th miniresonator.
Thus, as can be seen in Fig. \ref{m-th-atomic-excitaions}, atoms are excited in the $m$ miniresonator near the natural frequency of this miniresonator $\chi\Delta_{m}$  inside the spectral range $\chi\Gamma_{\Sigma}$ described by the Lorentzian form of \eqref{atomic_exit_m_resonator}.
Due to the interaction with the atomic ensemble, the  coefficient $\chi$ slightly shifts the eigen frequency of the $m$-th miniresonator.
The spectral width of excited atoms is close to the spectral interval between the nearest modes of miniresonators 
$\chi\Gamma_{\Sigma}\geq \frac{3\Delta}{\pi}\approx\Delta \gg\gamma_b$, significantly exceeding the eigen spectral width of the miniresonator mode. 
Thus, the  atoms excited in all the miniresonators completely overlap the spectrum of the signal pulse.

The expression \eqref{b-m-omega=2} reflects an important property of recording a signal pulse in a system of atoms in a miniresonator. 
Namely, no matter how small the intrinsic spectral width of the resonator $\gamma_b$, the signal is stored in a wider spectrum of atoms $\chi\Gamma_{a}^0\gg\gamma_b$, which  is determined by the collective interaction of atoms  (the spectral density of the atomic transitions and average coupling constant) with the field mode of $m$-th miniresonator.
An important conclusion follows from this: if, for example, between the common resonator and the miniresonators we even place another, narrower-band resonators,
then, its use will not allow us to narrow the spectral width of the signal pulse storage in the miniresonators, which will still be determined by a constant $\chi\Gamma_{\Sigma}\gg\gamma_b$.

By assuming a sufficiently narrow spectrum of a signal pulse $\delta\omega_s<\kappa/3$ and $\Gamma_{\Sigma}\gg\gamma_b$, we evaluate the transfer of this pulse to the all atomic ensembles ($t>\Gamma_{\Sigma}^{-1}$) by using Eq. \eqref{b-m-omega=2}:

\fla{
&P_a(t)=\sum_m^{M}\sum_j^{N_m}|S_{-}^{j,m} (t)|^2=
\nonumber \\
& \frac{2\pi M N}{\kappa} 
e^{- \frac{t}{T_1}}
\int  d\Delta_m
\frac{\delta_{in}}
{\pi(\delta_{in}^2+\Delta_{m}^2)}
 \cdot
\nonumber
\\
&\int d\delta_{j,m}
\frac{\Delta_{in}}
{\pi(\Delta_{in}^2+\delta_{j,m}^2)}
\frac{\abs{f_a g}^2 }
{\Gamma_{\Sigma}^2+
(\Delta_{m}-\frac{\delta_{j,m}}{\chi})^2}
|\tilde{A}_{in} (\delta_{j,m})|^2. 
\label{all-atoms}
}
Assuming a large inhomogeneous broadening of the atomic atomic transition, we have  $\frac{\Delta_{in}}
{(\Delta_{in}^2+\delta_{j,m}^2)}\cong \frac{1}{\Delta_{in}}$.
Next we calculate the integral $\int d\Delta_m ...$ when $\delta_{in}\pm\Gamma_{\Sigma}\gg\delta\omega_s$ that leads to

\fla{
P_a(t)\cong & e^{- \frac{t}{T_1}}\frac{2 M |g|^2}{\kappa(\delta_{in}+\Gamma_{\Sigma})} 
\frac{N\abs{f_a}^2 }
{\Delta_{in}\Gamma_{\Sigma}}
\int d\omega |\tilde{A}_{in}(\omega)|^2=
\nonumber \\
&e^{- \frac{t}{T_1}}
E_1 E_2 W_{in}.
\label{storage}
}
where $E_1=\frac{(\kappa-\gamma_c)}{\kappa}$, $E_2=\frac{\Gamma_a^0}{\Gamma_a^0+\gamma_b}$ and we use the impedance matching condition \eqref{impedance_match-small_Gamma}, $\Gamma_{\Sigma}=\Gamma_a^0+\gamma_b$ and $\Delta_{in}\gg\gamma_a$

The quantum storage efficiency  can be presented using the cooperativity parameter of the system under study $E=\frac{C}{C+1}$ as it has been shown in work \cite{gorshkov2007photon}.
In the QM studied, it can be done for $E_2= \frac{C_{mr}}{C_{mr}+1} $ where the cooperativity parameter $C=C_{mr}=\frac{\Gamma_a^0}{\gamma_b}$ corresponds to the field absorption in the miniresonator, where the total losses are determined by the absorption of the field mode by atomic ensembles with an absorption coefficient $\Gamma_a^0=\frac{N\abs{f_a}^2}{\Delta_{in}}$ (where the rate of coherence loss  is determined by the inhomogeneous broadening $\Delta_{in}$) and the intrinsic field losses in the miniresonator, determined by the decay constant $\gamma_b$.
Assuming that the intrinsic losses of the  common resonator are small ($\kappa/\gamma_c=C_{cr}\gg 1$), then with impedance matching,
the efficiency of signal pulse transmission into the common resonator is determined by the same relation
$E_1\cong \frac{C_{cr}}{C_{cr}+1}$,
where $\frac{\kappa}{2}+\frac{\kappa}{2}$ is the loss of the common resonator  into waveguide modes + miniresonator systems and $\gamma_c$ is the decay rate of the intrinsic losses.


As can be seen in the solution \eqref{storage}, the efficiency of signal transfer to the system of atoms is determined by two factors. 
The first factor $E_1$ describes the efficiency of signal transfer to the system of miniresonators.
Its efficiency in the presence of impedance matching is limited  by the losses in the common resonator ($\gamma_c$). 
The second factor $E_2$ is determined by the transfer of radiation from the miniresonator system to atoms and its efficiency depends on the radiation losses in the miniresonators $\gamma_b$. 
This consideration implies that at the loading stage, losses in the system by atoms can be neglected. 
Atomic losses become significant at subsequent stages of signal pulse storage. 
Thus the perfect transfer \eqref{storage} of a signal pulse   to a system of atoms occurs, as noted above, under the condition of impedance matching within given spectral range (see above Figs. 
\ref{Output_field-4-5}-\ref{Output_field-6-7}, as well as with sufficiently small losses in the common ($
\gamma_c\ll \kappa$) and in the miniresonators ($\gamma_b\ll \Gamma_a^0$).
Eq. \eqref{storage} also shows that further storage of the signal pulse energy in atoms is determined by the rate of longitudinal (energy) relaxation.
However, the retrieval of recorded information will be largely affected by the phase relaxation of atomic coherence at the stage of information storage in the system of atoms.

The analysis performed  completes the main analytical  study of the storage of a signal light pulse into atomic ensembles.
For long-term storage of recorded information, two laser $\pi$ pulses can be applied to atomic ensembles at an adjacent transition $\ket{2}\leftrightarrow\ket{3}$  to transfer the excited optical coherence of atoms to long-lived spin sublevels $\ket{1}\leftrightarrow\ket{2}$ (see Fig.\ref{Long-lived_storage}).
The subsequent retrieval of the stored signal is realized by rephasing the excited optical atomic coherence.
The recovering  (phasing) of optical atomic  coherence will start the  coherent interaction of atoms with the field mode in each miniresonator.
The radiation appearing in the miniresonators will be transmitted to the common resonator, and from it the radiation will exit into the waveguide in the form of a photon echo signal. 
The descriptions of these processes were carried out  numerically in works \cite{perminov2018superefficient,Haechan_APL2025} and showed the possibility of an efficient reconstruction of the light signal pulse.
However, many physical properties of the conditions for the effective implementation of this scheme remain unstudied.
Below, we briefly analyze the main properties of this process (analytical calculations are in many ways similar to those given above and are therefore given in the Appendix).

\section{The stage of signal retrieval}

In the implementation of photon-echo QM protocols, the restoration (rephasing) of atomic coherence is accomplished in various ways. 
These issues  determine the corresponding QM protocol and deserve special studies.
Here we will focus only on some of the issues that are of primary importance for the QM scheme under consideration, also noting issues that will require more detailed analysis in special studies (some technical details are also given in the Appendix).

\subsection{Dual CRIB-protocol} 
First of all, it is interesting to note the fundamental possibility of a theoretically ideal reversible scenario for restoring the input signal pulse, which we will call a Dual CRIB-protocol below. 
In this protocol, we invert the frequency detunings of both atoms and miniresonators to reconstruct the input signal in the echo pulse ($t>t_0$: $\Delta_{m}\rightarrow - \Delta_{m} $ and $\delta_{j,m}\rightarrow - \delta_{j,m} $).
It is noteworthy that in this case, the resulting dynamics of the interaction of atoms and miniresonators with the echo field emitted is described by a system of equations that is completely reversible with respect to the dynamics of signal pulse absorption when the impedance matching condition is met (see these equations in Appendix). 
Unlike the CRIB protocol in an optically thick medium \cite{moiseev2004photon}, in the case studied, no additional operations with atomic coherence are required to ensure phase matching, which somewhat simplifies its implementation.
In the Appendix, echo signal calculations are performed similarly to the absorption of the input signal pulse, differing in that at the initial time $t=t_0$, atomic coherence is specified in the atomic system. 
The calculations yield the following solution for the amplitude of the emitted echo signal as a time-reversed copy of the input signal pulse (see also Eq. \eqref{echo-final} in the Appendix):

\fla{
A_{echo,d-crib}(t)
\cong E_1 E_2 \mathcal{Q}(t_e) 
 e^{ -i\omega t_e}A_{in}(t_e-t),
\label{D-CRIB-echo-final}
}
where $\mathcal{Q}(t_0)=\exp\{-T_s/T_{2,1\leftrightarrow2}\} \exp\{-(t_0-T_s) \gamma_a \}$ is a factor that describes the influence of relaxation effects on atomic transitions $\ket{1}\leftrightarrow\ket{3}$ and 
$\ket{1}\leftrightarrow\ket{2}$ ($T_{2,1\leftrightarrow m}$ is a phase relaxation time on atomic transition $\ket{1}\leftrightarrow\ket{m}$),
$\tilde{b}_m(\delta_{j,m})$ is given by Eqs. \eqref{b-m-omega} and \eqref{a-omega}. 
From Eq. \eqref{D-CRIB-echo-final} we find that the efficiency of echo emission 

\fla{
E_{d-crib}= \left( E_1 E_2 \mathcal{Q}(t_e) \right)^2,
\label{efficiency}
}
which is equal to the efficiency of storing the signal pulse \eqref{storage} to the second power, is also affected by the decoherence of the atomic states at the signal pulse storage stage in the quantum memory cell.

Thus, under conditions of weak atomic relaxation and good quality resonators, the Dual CRIB-protocol provides theoretically 100\% efficiency. 
It is worth noting that the considered protocol is another option in the CRIB protocol family 
\cite{moiseev2004photon,kraus2006quantum,moiseev2013scalable} that provides an ideal variant of reversible dynamics. A notable feature of which is its insensitivity to the presence of strong spectral dispersion effects occurring during the resonant interaction of light in atomic ensembles. Due to this, for its effective operation, it is sufficient to ensure only the complete absorption of signal radiation by the atomic ensemble.

The experimental implementation of Dual CRIB-protocol, despite its apparent complexity, is entirely feasible, for example, in the microwave frequency range, where methods for dynamically controlling eigen frequencies of miniresonators  are developed in optics \cite{fang2022cavity-tunig} and microwave domain, where, it was recently applied for implementation of multiresonator QM protocol  \cite{2021-Bao-PRL}.
Methods for controlling atomic frequency detunings are already quite well developed and are often used in the implementation of the CRIB and AFC protocols \cite{Moiseev_fp2025}, see also recent work \cite{meng2025eff-nteg-QM}.
Further discussion of Dual CRIB-protocol implementation possibilities is given in Section IV.

Implementation of the Dual CRIB-protocol will currently require significant efforts to select the most suitable multiresonator scheme and experimental equipment to work with it.
In this regard, it is interesting to consider other possibilities.
Among these options, the QM protocols that exploit the natural nature of inhomogeneous line broadening are of interest \cite{moiseev2011photon,Damon2011, mcauslan2011photon,ma2021elimination},among them, we will focus on a ROSE protocol \cite{Damon2011} using 
Revival of Suppressed Echoes (ROSE) and its improved implementation -noiseless photon-echo (NLPE-) protocol \cite{ma2021elimination}, which provides stronger quantum noise suppression.

\subsection{ROSE protocol} 
\begin{figure}
\includegraphics[width=1.0\linewidth]{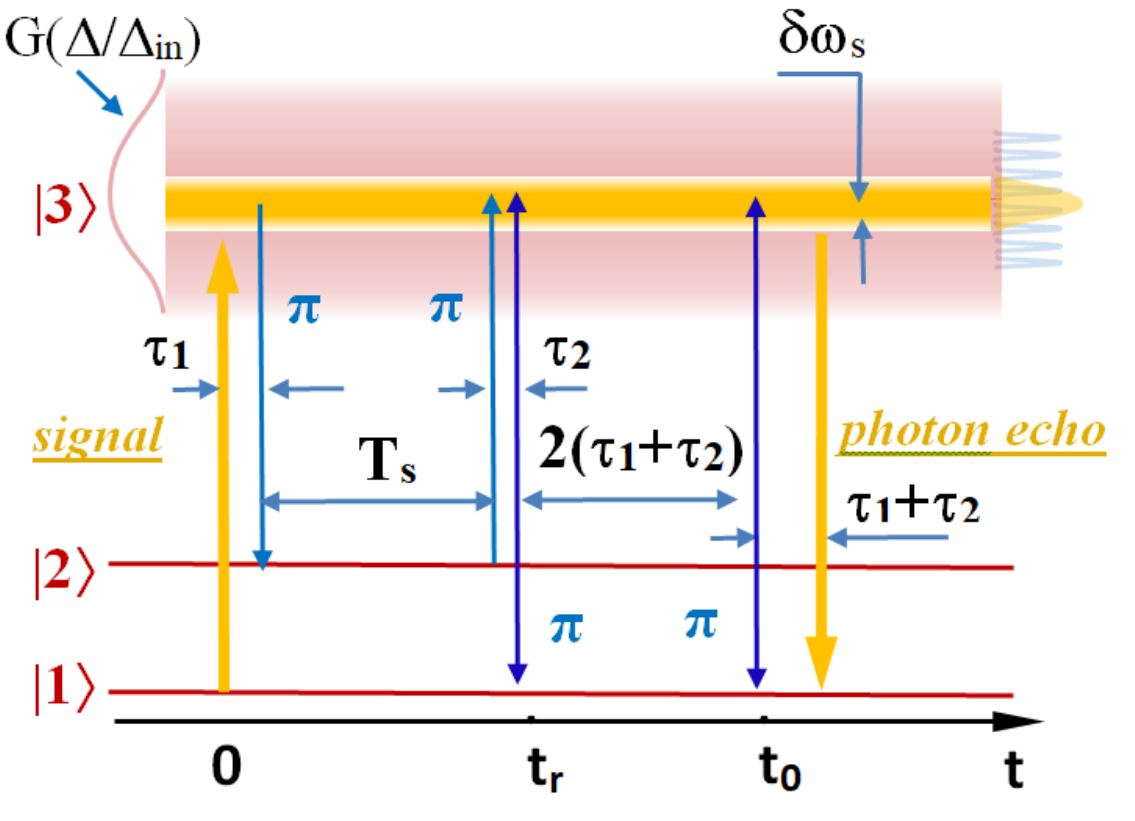}
\caption{\label{Long-lived_storage} 
A time sequence of a QM protocol with long-term storage of a signal pulse on an atomic coherence of the transition $\ket{1}\leftrightarrow\ket{2}$.
The optical transition $\ket{1}\leftrightarrow\ket{3}$ is inhomogeneously broadened, and the structure of the resonant (blue) lines of the miniresonators is shown on the right.  
After absorption of an signal pulse, the additional two laser $\pi$ pulses are applied to the adjacent auxiliary atomic transition $\ket{2}\leftrightarrow\ket{3}$  to transfer optical coherence to the long-lived transition $\ket{1}\leftrightarrow\ket{2}$ for the storage duration $T_s$.
At the transition $\ket{1}\leftrightarrow\ket{3}$, 2 more $\pi$ pulses are applied to restore the dephased atomic coherence, which then causes the emission of a photon echo after a delay time $\tau_1+\tau_2$. 
}
\end{figure}

This protocol (see Fig.\ref{Long-lived_storage})  uses two $\pi$ pulses that result in rephasing of the previously excited atomic coherence and restoration of the population inversion of the quantum transition at the moment of echo signal (ROSE-echo) emission.
However, in this case, it is also necessary to suppress the photon echo (primary echo) emission after the first $\pi$ pulse acting at the transition $\ket{1}\leftrightarrow\ket{3}$ or on the additional transition $\ket{1}\leftrightarrow\ket{4}$ \cite{ma2021elimination}, in which each of the atoms has the same frequency detunings as in some quantum transitions of  rare earth ions (see  also its recent implementation in the integrated scheme \cite{meng2025efficient}).
Primary echo suppression can be achieved by breaking phase-matching  and/or introducing additional dephasing of atoms.
The latter is successfully implemented for the CRIB, GEM and AFC protocols by applying controlled gradients of electric, or magnetic fields, causing additional shifts in atomic levels due to the Zeeman and Stark effects (see, for example,  recent review \cite{Moiseev_fp2025}). 
\noindent
It can be added here that the use of compact miniresonators, including planar ones, with atomic ensembles placed near them, appears to be a convenient platform for local control of external magnetic and electric fields on atoms and excitation of the atomic ensemble by evanescent fields of laser pulses (see, for example, work \cite{Sellars2015}).

Moreover, AFC- and  ROSE-protocols  do not provide perfect time reversibility, which can lead to a negative effect of spectral dispersion on the efficiency and fidelity in echo signal emission. 
In our case, spectral dispersion manifests itself in the dependence of the atomic coherence in $m$-th miniresonator
$S_{j,m}\sim \left ( \chi\Gamma_{\Sigma}+i(\chi\Delta_{m}-\omega_{j,m})\right)^{-1} \tilde{A}_{in} (\omega)$ excited during the absorption of a signal pulse 
(see Eqs.\eqref{atom-coherence-2}, \eqref{b-m-omega}).
When implementing the ROSE protocol,
after two short laser pulses are applied to all miniresonators,
the light field emitted by the atoms of the $m$-th resonator into the common resonator (and into free space) experiences the same spectral dispersion effect $\sim \left( \chi\Gamma_{\Sigma}+i(\chi\Delta_{m}-\omega)\right)^{-2} \tilde{A}_{in} (\omega)$, which leads to a strong suppression of the photon echo emission efficiency (see Appendix).

We show that the two problems noted above: 1) suppression of the primary echo emission and 2) strong attenuation of the spectral dispersion in the ROSE-echo retrieval  - can be solved for a sufficiently large spectral width of the miniresonator system compared to the spectral width of the signal pulse ($
\delta_{in}\gg\delta\omega_s$). 

1) To suppress the primary echo, the first  laser $\pi$ pulses can be used, exciting the nearest min-resonators with phases $\varphi_{1;m}$ that differ by $\pi$.
With an even number of miniresonators, their frequencies must be symmetrically arranged relative to the frequency of the common resonator ((for example $\varphi_{1;m}=\varphi_m=(\frac{1}{2}+m)\pi$, $m=\frac{M}{2}, ...,-2,-1,0,1,..., \frac{M}{2}-1$)). 
In this case, the phasing polarization corresponding to the radiation of the primary echo by all miniresonators will be completely out of phase due to the anti-symmetry of the state of atomic coherence of the whole and therefore will not cause the appearance of a primary echo. 

2) To reduce the effect of spectral dispersion, we choose the phases $\varphi_{2;m}$ of the second $\pi$  laser pulses to be the sum of two terms: $\varphi_{2;m}=\delta\varphi_{2;m}^{(1)}+\delta\varphi_{2;m}^{(2)}$, where $\delta\varphi_{2;m}^{(1)}=\varphi_{1;m}+\pi$, and the other is equal to $\delta\varphi_{2;m}^{(2)}=-2\arctan\{\frac{\Delta_m}{\Gamma_{\Sigma}}\}$ (see Appendix).
The phases $\delta\varphi_{2;m}^{(1)}$ align the atomic phases in nearby miniresonators, and the phases $\delta\varphi_{2;m}^{(2)}$ neutralize the influence of spectral dispersion, so that the overall behavior of atomic coherence becomes similar to the Dual CRIB-protocol  under the condition $
\delta_{in}\gg\delta\omega_s$ and leads to 

\fla{
A_{Rose-echo}(t)
\cong E_1 E_2 \mathcal{Q}(t_e) 
 e^{ -i\omega t_e}A_{in}(t-t_e),
\label{Rose-echo-final}
}
with the same efficiency of the photon echo emission as in 
\eqref{efficiency}.

Thus, using a large number of miniresonators opens the possibility of suppressing spectral dispersion, unlike implementing ROSE-protocol (and AFC-protocol) in a single-resonator scheme \cite{Afzelius-PRA-2010,Sabooni2013,2021-PRB-Minnegaliev}.
At the same time, it should be noted that the proposed spectral dispersion compensation leaves the signal's temporal shape unchanged, which still reproduces the temporal shape of the signal pulse, which is similar to the spectral dispersion compensation methods considered in the works  \cite{moiseev2012rephasing,arslanov2017optimal,EMoiseev2021}.
The experimental implementation possibilities of the ROSE- protocol are discussed in Section IV.

\subsection{Quantum noises}. 
Studies conducted by several authors have demonstrated the need for quantum noise suppression in the ROSE protocol (see, for example, reviews \cite{Chaneliere2018,Moiseev_fp2025}). Noise arises from the presence of a large number of excited atoms at the moment of echo signal emission, which can be caused by imprecise implementation of laser $\pi$ pulses, as well as the influence of strong relaxation of atoms from the excited state to the ground state after the first control laser $\pi$ pulse.

To estimate the level of emerging quantum noise, in Appendix, calculations were made of the number of photons spontaneously emitted by atoms into the common resonator at the stage where the depopulation of the resonant energy levels involved in this relaxation process can still be neglected.
A detailed analysis of optical quantum noise was carried out for the following parameters of the multiresonator circuit under consideration:
$\Delta_{in},\delta_{in}\gg\kappa\gg\Gamma_{a}^0$. 
Choosing other parameters likely will not change the overall behavior significantly. Determining the precise kinetics of spontaneous photon generation requires numerical calculations, which will be of interest when analyzing a specific experimental situation.
The development of a more general theory is the subject of a separate discussion. 

The average number of photons spontaneously emitted into the common resonator $n_{r}(t)=\langle \hat{a}^{\dagger}(t)\hat{a}(t)\rangle$ is described by the solution (see  derivation of Eq. \eqref{aa-noises-3} in the Appendix):

\fla{
n_{r}(t)&\cong \frac{ \Gamma_a^0 \kappa}{\frac{\kappa}{2}+\ae(t_r)-\Gamma_a(t_r)} P_{33}(t_r)\cdot
\nonumber
\\
& \cdot  
\bigl\{\frac{1-e^{-2\Gamma_a(t_r)(t-t_r)}}{\Gamma_a(t_r)}-
\frac{1-e^{-2(\frac{\kappa}{2}+\ae(t_r))(t-t_r)}}
{\frac{\kappa}{2}+\ae(t_r)}
\bigr\},
\label{aa-noises-general}
}
where $\ae(t_r)=\frac{M|g|^2}{\delta_{in}+\Gamma_a(t_r)}$, $\Gamma_a(t_r)=-w(t_r)\frac{N_a f_a^2}{\Delta_{in,a}}$ ,
$w(t_r)=\langle \hat{w}\rangle_{t_r}$ is an average inversion on the resonant atomic transition 
$\hat{w}_j =\ket{3_j}\bra{3_j}-\ket{1_j}\bra{1_j}$.

Of interest is the behavior of the number of photons over sufficiently long times, when the condition $\kappa (t-t_r)\gg 1$ is satisfied, where $t_r$ is the moment in time from which the evolution of the considered system of atoms and resonators begins after excitation in mini resonators, for example, after exposure to the first laser $\pi$ pulse (see Fig. \ref{Long-lived_storage}).
In the studied parameters of the interaction we have 
$|\Gamma_a(t_r)|\ll \kappa $ and  
$\ae(t_r)=\frac{M|g|^2}{\delta_{in}+\Gamma_a(t_r)}
\cong\frac{M|g|^2}{\delta_{in}}\cong\kappa/2$), 
that transforms Eq. \eqref{aa-noises-general} into the following simple expression for the number of photons in a common resonator:

\fla{
n_{r}(t)\approx \Gamma_a^0   P_{33}(t_r)\cdot
\bigl\{\frac{1-e^{-2\Gamma_a(t_r)(t-t_r)}}{\Gamma_a(t_r)}-\frac{1}{\kappa}
\bigr\}.
\label{aa-noises-5}
}

It is noteworthy that the number of noise photons in the common resonator is determined by the probability of excitation of an individual atom $P_{33}(t_r)$, and not of an atomic ensemble of atoms at $|2\Gamma_a(t_r)(t-t_r)|\leq 1$, which is due to the properties of the resonator assistant QM, operating under impedance matching conditions.
In the case where excited levels are predominantly populated, when $w(t_r)>0$ and 
$\Gamma_a(t_r)=-w(t_r)\frac{N_a f_a^2}{\Delta_{in,a}}=-w(t_r)\Gamma_a^0 < 0$,  
we get an exponentially increasing rate of spontaneous generation (increased luminescence) of photons:

\fla{
n'_t(t)=2\Gamma_a^0\exp\{2\Gamma_a^{0}w(t_r)(t-t_r)\} P_{33}(t_r).
\label{photon-generation}
}

The solutions \eqref{aa-noises-5}, \eqref{photon-generation} remain valid as long as the population of the excited level $N_{2}(t_r)$ changes negligibly, that is, the condition is satisfied 
$ \frac{ P_{33}(t_r) }{|\omega(t_r)|} \exp\{2\Gamma_a^{0}w(t_r)(t-t_r)\}\ll  N_a w(t_r)\frac{\kappa}{\Delta_{in,a}}=N_{2}(t_r) $. 
We took into account that the number of participating atoms  in each mini resonator is approximately equal $N_a \frac{\Gamma_a^0}{\Delta_{in.a}}$, while the  total number operating miin-resonators   $N_r=\frac{\kappa}{\Gamma_a^0}$ (for $\delta_{in}\gg\kappa$, $\frac{\kappa}{\Delta_{in,a}}\approx 0.01$).
At large numbers of atoms $N_a$ and  $w(t_r)\approx 1$, we get $N_{2}(t_r)\gg1$. 
In this case, it is quite possible to use a sufficiently long time to ensure coherent control of atoms $(t-t_r)\gg (2\Gamma_a^0)^{-1}$ without significantly affecting the total number of atoms in the excited state.
This occurs because the atomic system has a sufficiently large inhomogeneous broadening and there are many miniresonators with different frequencies, which also suppresses the  coherent amplification in photon  emission.

We can apply the solution \eqref{aa-noises-5} to the photon echo signal readout stage. In this case, the initial time in the solution is $t=t_0$ (see Fig. \ref{Long-lived_storage}).
The population of excited level is small ($P_{33}(t_0)\ll 1$), the atomic inversion becomes negative $w(t_0)<1$ and close to $-1$ and parameter $\Gamma_a (t_a)\approx \Gamma_a^0$. 
In this case, the solution \eqref{aa-noises-5}  shows that spontaneous generation of noise photons reaches a steady state (for $2\Gamma_a(t_0)(t-t_0)>1$):

\fla{
n_{r}(t)&\approx \Gamma_a^0   P_{33}(t_0)\cdot
\bigl\{\frac{1-e^{-2\Gamma_a(t_0)(t-t_0)}}{\Gamma_a(t_0)}-\frac{1}{\kappa}
\bigr\}\approx 
\nonumber
\\
&\frac{P_{33}(t_0)}{|w(t_0)|}\ll 1.
\label{aa-noises-4}
}

The number of noise photons increases strongly as the atomic transition approaches saturation $w(t_0)\rightarrow 0$.
With a low average  population of the excited transition $P_{33}(t_0)\ll1$, the number of photons in the common resonator becomes very small and slowly changing. 
Taking into account the dependence of the light field operator in the waveguide on the field in the resonator ($\hat{a}(t)=\sqrt{\kappa}\hat{a}_r (t)$), we find the number of emitted noise photons measured by the detector within a time interval equal to the duration $\delta t_s\approx \delta\omega_s^{-1}$ of the photon echo signal 

\fla{
n_{noise}(t_e)=\kappa \int_{t_e-
\delta t_s/2}^{t_e+
\delta t_s/2} dt \langle \hat{n}_r(t) \rangle = \frac{\kappa
}{\delta\omega_s } \frac{ P_{33}(t_0)
}{w(t_0)}.
\label{detected-noise}
}

By imposing a condition on the maximum allowable number of noise photons $n_{n,max}$ in the photon echo signal,
we find the requirement for the maximum allowable excitation probability of an atom to the upper level at the moment of an echo signal emission:

\fla{
P_{33}(t_0)\leq \frac{\delta\omega_s }{\kappa
}  \frac{n_{n,max}}{1+2\frac{\delta\omega_s }{\kappa
} n_{n,max} }\cong \frac{\delta\omega_s }{\kappa
} n_{n,max}.
}

Assuming $ n_{n,max}=0.01$ and $\frac{\delta\omega_s }{\kappa}\approx 0.3$ \cite{Moiseev2010cavity,EMoiseev2021},  we get a quite high requirement for depopulation of the excited level $P_{33}(t_0)\leq 0.003$.
Achieving such low photon noise levels seems to be feasible using the NLPE protocol.
It is worth noting that the implemented scheme of the NLPE protocol requires suppressing quantum noise arising from the luminescence of atoms at adjacent optical transitions. 
When implementing this protocol in resonator and multiresonator schemes, the noise photons  can be suppressed by its filtering  using the resonators of the QM circuit itself.
Understanding how this can be implemented in a real situation requires a special study that takes into account the spectroscopic parameters of the atoms used.

\section{issues of experimental implementation}

Here, we discuss the possible experimental implementation of the QM protocols described. 
Similar to the development of existing QM  protocols based on photon echo \cite{Moiseev_fp2025}, the main challenges arise in finding ways to restore atomic coherence that ensure high-quality photon echo emission. 
In our case, signal pulse retrieval requires controlling the coherence of atomic ensembles located in several compact high-Q miniresonators.
Below I briefly describe possible ways to implement the proposed QM  protocols using existing integrated technologies for the fabrication of high-Q miniresonators and methods for controlling its parameters and atomic ensembles.

In the last 10 years, there has been a rapid development of integrated photonic platforms, which have opened the possibility of creating various compact resonators with a high Q-factor, that meet the requirements for use in quantum technologies,
see, for example, recent reviews \cite{pelucchi2022potential, xie2024recent,labbe2025thin,ye2023foundry,liu2025recent}.
For the proposed QM, the following variants of high-Q miniresonators may be of interest:
small-sized ring resonators with whispering gallery light modes   \cite{gao2021broadband,nijem2024high,Lau2025},  nano-photonic resonators \cite{zhong2017nanophot,lukin2020integrated}, resonators fabricated  in nanostructured materials \cite{merkel2020-nano-struct} using silicon microcavity arrays,   
\cite{wachter2019silicon},
on-chip high-Q microresonators  on thin-film lithium niobate 
\cite{zhu2024twenty},
integrated $Si_3 N_4$ bus-coupled ring-resonators \cite{puckett2021422}.
Among the best results, we note the demonstrated microring resonators on lithium-niobate-on-insulator (LNOI) wafer with ultra-high Q-factors  $4\cdot 10^7$ \cite{li2023ultra}
 and $1.23\cdot 10^8$ \cite{gao2022lithium}
on $\lambda=1.55\mu$m with a resonator linewidth close and even  less than 1 MHz.
Such spectral parameters of the microring resonator correspond to the requirements of the proposed QM based on the use of high-Q miniresonators. 
Moreover, the use of LNOI-platform also makes it possible to tune the resonator frequency using compact electro-optical (EO) switchers \cite{savchenkov2003tunable}, which expands the possibilities of practical applications of such resonators.
Of particular note are the results of recent work \cite{barya2025ultra} that demonstrate the possibility of achieving ultra-high Q-factor  $\geq 10^{8}$  in ring resonator  on LNOI doped by erbium ions  with EO-switcher due to  the propagation of the light field in a highly dispersive ring medium.
Importantly, the use of erbium ions is essential for creating optical QM in quantum communications. 
It was recently demonstrated that erbium ions can be implanted into lithium niobate without critically degrading its spectroscopic properties \cite{wang2020incorporation}, opening the possibility of implementing QM  at the telecommunications wavelength $1.5 \mu$m on the LNOI platform.
We also note  recent works that successfully demonstrated a photon echo QM protocol on rare-earth ions doped in thin-film lithium niobate \cite{dutta2023atomic} and in nanophotonic resonator fabricated on the LNOI platform \cite{barya2024towards}.

The considered multi-resonator  Dual-CRIB and ROSE protocols are based on different types of inhomogeneous broadening of resonant atomic lines and methods of controlling the atomic coherence excited in them, ensuring efficient retrieval of the stored light field.
We will primarily  focus on the ROZE protocol (see Section III,B), in which atomic coherence control is based solely on the interaction of atoms with two laser control $\pi$-pulses. 
As discussed in Section III.B (see also Appendix C), each pair of laser pulses (with their specific phases) is fed independently  to the atoms of a selected QM miniresonator, eliminating the effect of these pulses on the atoms of other QM miniresonators.
In addition, ensuring control of atomic coherence by each pair of laser pulses should not worsen the storage of a signal pulse on the atomic ensemble states and the subsequent its retrieval in the photon echo.

\textbf{ROSE protocol}. Here we show how the atomic coherence control described in Sec.III can be achieved by creating the multi-resonator QM scheme on LNOI platform with the addition of EO switchers of resonator frequencies. 
It should be noted that, at present, the technology for creating such switches is quite well developed (see references \cite{wang2018integrated,li2020lithium,azeem2021dielectric,xia2022tunable}). 
In existing experiments, this method allows for fairly rapid frequency shifts of resonators by gigahertz while maintaining a high Q-factor.

\begin{figure}
\includegraphics[width=0.8\linewidth]{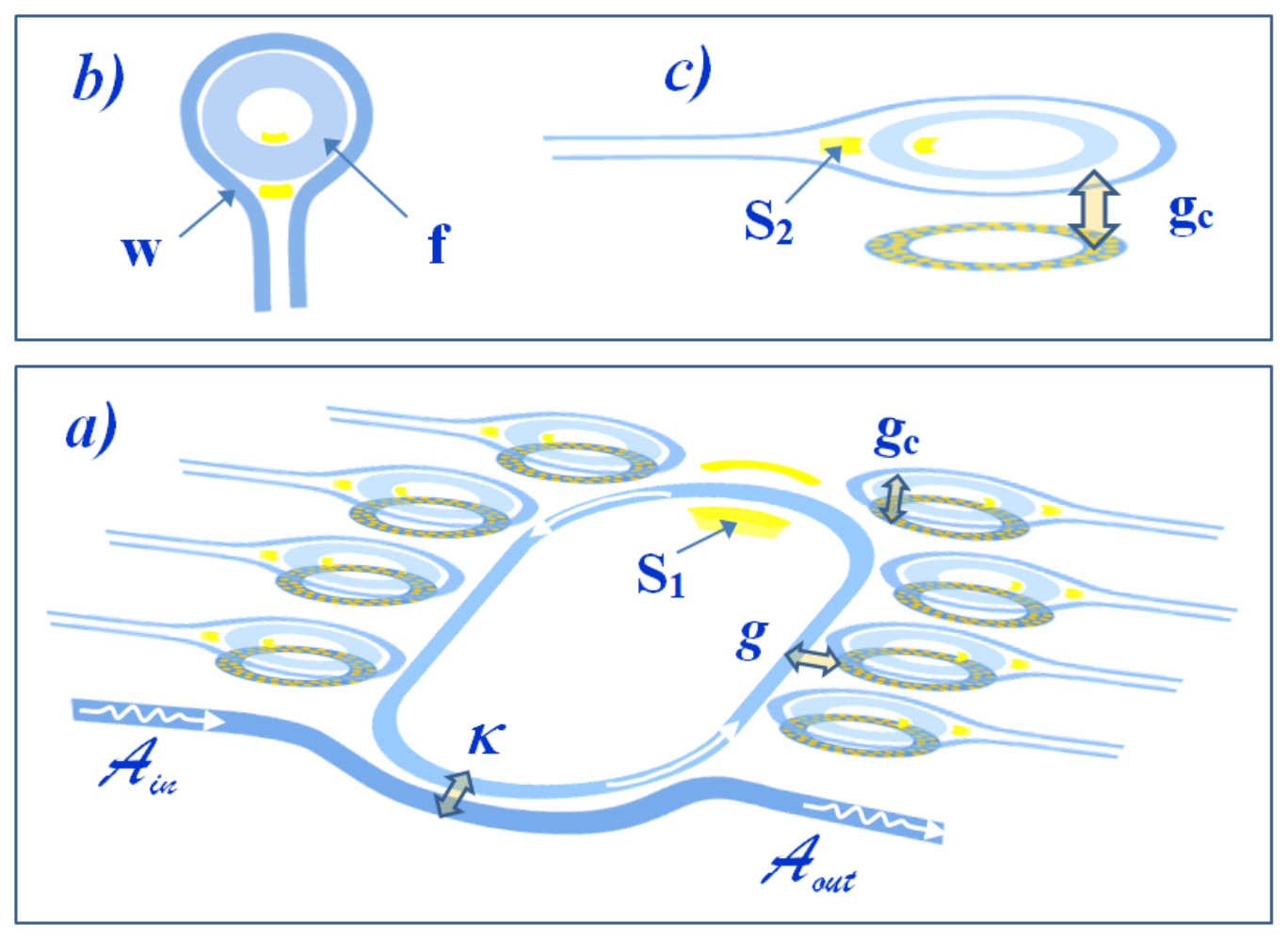}
\caption{\label{QM_scheme-experiment} 
a) possible  experimental scheme using system of ring miniresonators on LNOI platform; electro-optical (EO) switchers $S_1$ and $S_2$ (two yellow blocks) regulates the carrier frequencies of the common resonator and additional (feeder-"f") ring miniresonators; $g$ and $\kappa$ are the coupling constants of common waveguide with external waveguide and QM miniresonators.
b) an additional external m-th waveguide $w$ coupled with m-th f"- miniresonator  through which the radiation of the control laser pulses enters m-th QM miniresonator and excites the atoms located in it; 
c) the coupling (with the coupling constant $g_c$) of the f-miniresonator and m-th QM miniresonator is regulated by the EO switcher $S_2$  (yellow blocks), which can change the frequency of the f-miniresonator, causing the presence or absence of resonant interaction between these two resonators. 
}
\end{figure}

The proposed QM scheme is shown in Fig. \ref{QM_scheme-experiment}, where it can be seen that the basic QM scheme shown in Fig. \ref{QM_scheme} is supplemented with EO switchers $S_1$ and $S_2$ (shown as yellow blocks) that can change the frequencies of the main resonator and of the feeder ("f") miniresonators at certain periods of time.
It is worth noting that according to the theory presented and recent work \cite{matanin2025}, the common resonator can have a much lower Q-factor compared to the Q-factor of miniresonators, as long as it allows efficient transfer of the signal pulse from the common resonator to the miniresonators.
Furthermore, the frequency of the common resonator will change only when that resonator is free from signal radiation.
Therefore, the possible appearance of quantum noise during the activation of the EO switcher $S_1$ will not directly lead to the appearance of parasitic photons in atomic ensembles due to their rapid leakage into the external waveguide.

Before exciting the atoms in each QM miniresonator with control laser pulses, EO switcher $S_1$ significantly shifts the common resonator frequency by a sufficiently large detuning:
$\Delta_{sh}\gg \delta\omega_s$ ($\delta\omega_s$ is a spectral width of the signal pulse), putting it out of resonance with each QM miniresonator. 
Also immediately before exciting the atoms with an arbitrary control laser pulse, EO switcher $S_2$ tunes the frequency of the f-miniresonator to the frequency of the QM miniresonator.
Under these conditions, laser radiation emerging in each QM miniresonator will virtually never escape into the common resonator and, accordingly, will not excite atoms in neighboring QM miniresonators.
However, during the signal pulse storage stage, the f-miniresonator frequencies should be significantly detuned from the QM miniresonator frequencies, preventing significant degradation of the QM miniresonator's intrinsic Q factor.
This frequency mismatch should significantly exceed both the interaction constant $\Delta_{sh}\gg g_c \gg \gamma_b$ and the spectral width of the signal pulse, $\Delta_{sh}\gg \delta\omega_s$.
In this case, the QM miniresonator frequency shift becomes quite small ($\frac{|g_c|^2}{\Delta_{sh}} \ll \Delta_{sh}$) ($g_c$ is the coupling constant between the QM miniresonator and the f-miniresonator), without significantly affecting the structure of the eigen frequencies of the multi-resonator QM.
In addition to the above discussion of the ROSE protocol, it should be noted that the use of EO modulators facilitates the control of the phases of the control laser pulses.

The implementation of NLPE protocol (see Sec.III and Appendix C ), compared to the ROSE protocol, involves the use of two additional laser pulses with different frequencies.
In this protocol, the EO switchers $S_2$ must provide a larger  spectral shift of the f-miniresonators, which is possible due to the ability of such switchers to change frequencies.
In case of insufficiently strong evanscent field of the control laser pulse, it will also be necessary to tune  the frequency of the corresponding  QM miniresonator by incorporating its own EO switcher into its circuit.
It is important that changing the frequency of miniresonators, accompanied by a decrease in their Q-factor, will improve the implementation of the ROSE and NLPE protocols, since the exciting intense laser field will leave the miniresonators faster.

\textbf{Dual CRIB protocol}. 
Here one can describe the basic procedure for implementing this protocol  using the properties of the multi-resonator QM scheme discussed above.
In the CRIB protocol, all the QM miniresonators used contain compact EO switchers (similar to  switcher $S_2$, see above) that allow for inverting frequency offsets of these miniresonators during the signal retrieval in the photon echo. 
In addition, it is also necessary to be able to invert the frequency detunings of atoms. 
In known experimental schemes for implementing the CRIB/GEM protocol, this procedure is accomplished using switchable gradients of electric, or magnetic fields \cite{alexander2006photon,hosseini2011high,Moiseev_fp2025}. 
In the QM scheme under study, this procedure can also be realized by taking advantage of the small size of not only the QM miniresonators but also the entire circuit.
The small size of the entire QM circuit allows all the QM miniresonators to be placed within a common gradient field. 
When using EO switchers that use electrical current, one can use a magnetic field gradient, since such a field will not affect the operation of the switchers.
The gradient field remains constant during the signal pulse storage.
To retrieve the signal pulse in the photon echo, the gradient field must be inverted. 
It's worth noting that before applying the gradient field, a narrow line is prepared in the atomic system at the optical \cite{alexander2006photon}, or Raman \cite{hosseini2011high} transition. 
This is typically accomplished using the laser hole burning technique \cite{nilsson2004hole}, which is particularly widely applied in AFC protocol implementation.

Thus, the proposed multiresonator QM can be implemented on the integrated LNOI platform.
At the same time, the methods described indicate challenging tasks that need to be solved in order to achieve effective experimental implementation of the studied protocols.
Among the main challenges is the need to integrate multiple devices on a single platform, including the creation of sufficiently high-Q coupled QM miniresonators, f-miniresonators and switchers capable of precisely controlling the frequencies of these miniresonators. 
At the same time, this QM enables control of the quantum state of a stored light field by using additional manipulations of the atomic and field states in each QM miniresonator. 
Thus, the complexity of the multiresonator QM design enables both the storage of wider-bandwidth signal pulses and expanded functionality for quantum processing with stored quantum states. 
Therefore, it is expected that this QM will find widespread applications in quantum computing.


\section{Discussion and conclusion}

A theory of hybrid QM on atomic ensembles in a multi-resonance scheme, including a common resonator and several miniresonators containing atoms, has been developed.
The analytical results obtained on its basis allowed to establish a physical picture of the process  of a signal pulse storage  on a system of atomic ensembles.
Analytical relations determining the conditions for the effective implementation of the considered QM were obtained.
The spectral properties of the QM scheme under consideration were studied in detail.
It has been established that effective quantum storage  should be accompanied by the excitation  of atomic coherence in each of the miniresonators in a sufficiently large spectral range $\Gamma_a^0$, which is determined by the collective interaction of atoms with the miniresonator mode.
The spectral range $\Gamma_a^0$ should not be less than the frequency interval between the frequencies of the nearest miniresonators, exceeding the spectrum width of the eigenmode of the miniresonators.
The studied cascade multiresonator QM scheme can provide a significant reduction in the number of atoms required for use in miniresonators, 
which opens up the additional possibility of weakening the decoherent effects in the system of atoms.
This effect  will be significantly enhanced when using small-sized high-Q miniresonators. 
It is shown with significant suppression of spectral dispersion arising from the use of multiple miniresonators. This can be achieved using the Dual-CRIB protocol and the ROSE protocol (and in NLPE-protocol - its improved version). 
The latter two protocols require the use of appropriate phases for the laser fields used to excite atoms in different miniresonators.

The obtained analytical relation allows us to determine the optimal physical parameters of the considered QM.  
An impedance matching condition has been found that ensures high efficiency of the memory under consideration, and it has been shown that the presence of atomic ensembles can significantly influence this condition.
In this case, a spectral matching condition has been found, the fulfillment of which plays a major role in increasing the working spectral width of quantum memory.
It is also shown that the enhancement of the interaction of atoms with miniresonators strongly affects the impedance matching condition and it is found that the interaction enhancement  can also lead to a significant broadening of the working spectral range of the memory.
Based on the analysis carried out, conditions were also found under which a sufficiently strong suppression of quantum noise is possible when emitting an echo signal in the studied quantum memory scheme.

Fundamentally new possibilities in the control of optical QM appear due to the creation of a long-lived macroscopic coherence in it \cite{Moiseev_PRL2025}. 
The implementation of such a QM in a multiresonator scheme is of particular interest, especially for the efficient quantum conversion of the photon frequency.
Among potentially interesting atomic ensembles, we note rare earth ions in crystals \cite{GOLDNER-2015,Guo2023, Moiseev_fp2025}.
By choosing one or another rare earth ion, it is possible to obtain various resonant frequencies, including working at a wavelength of 1.5 $\mu$m, used in optical commercial lines. 
The recently demonstrated implementation of nano photonic crystals with rare earth erbium ions \cite{Yong2023PRL} opens up promising prospects for the implementation of the proposed QM.

In conclusion, we note that in the multiresonator QM  under study, the quantum state of the input signal pulse is stored in the form of an entangled quantum state of spatially separated atomic ensembles.
In addition, the use of a small number of miniresonators opens up new physical properties and possibilities for achieving highly efficient efficient QM \cite{Perminov2023} due to topological changes in the spectrum of its eigenmodes.
Here, multiresonator QM containing single atoms \cite{moiseev2022multiresonator}  is of particular interest, since it provides additional opportunities for the implementation of quantum gates on atomic qubits.
Thus, the system of individual atoms and/or atomic ensembles, connected to each other, form a controlled  artificial molecular system of significant size (tens to hundreds of $\mu$m) due to the strong coupling of miniresonators and the long lifetime of atomic coherence.

Thus, experimental development of the memory under study is already possible using existing integrated optical technologies on LNOI platform. 
The proposed multiresonator QM can also be implemented using a system of superconducting microwave resonators, for which methods for controlling the frequencies of such resonators and their interactions have already been reliably demonstrated and used. 
Further experimental research is needed to study such multiresonator systems containing spin ensembles.
The methods for implementing these  QMs  can be complemented by various scenarios for controlling the quantum dynamics of the atoms (spins) in a QM cell, opening the possibility of implementing basic  protocols of quantum storage and processing with a system of photonic qubits loaded into it. 
Studying these issues opens up independent further research.

\begin{acknowledgments}
The author thanks N.S.Perminov, K.I.Gerasimov and  M.M.Minnegaliev for useful discussions.
This research was supported by the Ministry of Science and Higher Education of the Russian Federation (Reg. number NIOKTR 125012300688-6).

 \end{acknowledgments}

\section{Appendix: the light field retrieval}

\subsection{Spectral properties of quantum dynamics}

Using the Fourier transform for the mode amplitudes ${b}_{m} {(t)}=\frac{1}{\sqrt{2\pi}}\int d\omega \tilde{b}_{m} {(\omega)}e^{-i\omega t} $ and $a(t)=\frac{1}{\sqrt{2\pi}}\int d\omega \tilde{a} {(\omega)}e^{-i\omega t} $, we get the solution of Eq.\eqref{eqfa-5}:

\fla{
 \tilde{b}_{m}{(\omega)}= &
-i g_m^{*}\frac { 1}
{\gamma_b +\frac{N_m |f_m|^2 }{\Delta_{in,a}-i\omega}+i(\Delta_m-\omega)}\tilde{a} {(\omega)}=
\nonumber \\
&-i g_m^{*} \mathcal{R}_m(\Delta_m,\omega)\tilde{a} {(\omega)},
\label{eqfa-5-dop-copy}
}
where $\Delta_{in,a}=\Delta_{in}+\gamma_a$.

For simplicity of the experimental implementation of the quantum memory scheme under consideration, we also assume that spectral width of signal pulses is much smaller than the inhomogeneous broadening of the atomic optical transition $\delta\omega_s\ll\Delta_{in}$. In this case

\fla{
\frac{N_m |f_m|^2 }{\Delta_{in,a}-i\omega}
&\cong \Gamma_{a,m}^0 (1+i \frac{\omega }{\Delta_{in,a}})=
\nonumber \\
&\Gamma_{a,m}^0+i\tilde{\chi} \omega,
\label{approx-Ap}
}
where $\tilde{\chi}=\frac{N_m f_m^2 }{\Delta_{in,a}^2}\ll 1$; 
$\Gamma_{a,m}^0=\frac{N_m |f_m|^2}{\Delta_{in,a}}$ is  the relaxation constant, which  is a result of collective interaction of atoms with the resonator mode and determines the absorption rate of the $m$-th resonator mode by the atoms of this resonator. 
As shown below, this parameter also determines the spectrum width within which an ensemble of atoms with a large inhomogeneous broadening of the resonant transition is excited (see also Fig.\ref{m-th-atomic-excitaions} below).

Using the approximation \eqref{approx-Ap} in $\mathcal{R}_m(\Delta_m-\omega)$, we obtain

\fla{
\mathcal{R}_m(\Delta_m,\omega)\cong 
\frac { \chi}
{\chi(\gamma_b +\Gamma_{a,m}^0)+i(\chi \Delta_m-\omega)},
}
where $\chi=\frac{1}{1-\tilde{\chi}} \cong 1+\tilde{\chi}$. 

Thus, the interaction with atoms shifts the miniresonator line ($\Delta_m\rightarrow\chi\Delta_m$), increases its linewidth ($\gamma_b\rightarrow\chi(\gamma_b+\Gamma_{a,m}^0)$) and coupling constant with a common resonator ($g_m\rightarrow\chi g_m$).
Taking into account large inhomogeneous broadening ($\Delta_{in}\delta t_s\gg 1$, where $\delta t_s$ is a temporal duration of a signal pulse) in Eqs.\eqref{eqfa-5} and \eqref{eqfa-5-dop-copy}, we can write a solution for $b_m (t)$ in the form: 

\fla{
 b_{m} ( t)& \cong - i g_m^{*} \chi
  \int_{-\infty}^t dt' \cdot
  \nonumber
  \\
 &\exp\{-\chi\left( i\Delta_{m}+\gamma_b + \Gamma_{a,m}^0 \right) (t-t')\}{a}(t').
 \label{b-m-amplitude-AP}
}

The large inhomogeneous broadening of the atomic transition ($\Delta_{in}\gg\delta_{in}$ where $\delta_{in}$ is a linewidth of multi-resonantor system, see Fig. 2) allows using the same atomic ensembles in each  miniresonator.
Assuming the atomic ensembles in each of the resonators to be identical (i.e. $\Gamma_{a,m}^0=\Gamma_{a}^0$),
we will have for the total attenuation constant of each miniresonator $\Gamma_{a,m}^0+\gamma_b\equiv
\Gamma_{a}^0+\gamma_b=\Gamma_{\Sigma}$ ($\Gamma_{a}^0=\frac{N_a f_a^2}{\Delta_{in,a}}$, $N_a$ is a number of atoms in each miniresonator $f_a$ is an average coupling constant).

In carrying out further calculations, we will focus on studying the temporal dynamics at times shorter than the time of automatic rephasing of miniresonators $
\tau=\frac{2\pi}{\Delta}$ . 
In such short time intervals ($t\ll\tau$) of the behavior of the system under consideration, it is possible to switch from a discrete frequency distribution $G_{r} (\Delta_m/\delta_{in})$ of resonators to their continuous distribution $\tilde{G}_{r} (\Delta_m/\delta_{in})$.
(see, for example,  mathematical description of similar transition in  \cite{moiseev2012rephasing}).
Accordingly, bearing in mind the finding of the field of the common resonator $a(t)$ at these time intervals, in the expression \eqref{a-omega}, we replace the sum with the integral over the frequencies of the mini resonators:  
\fla{
&\int d\Delta_m  G_{r} (\frac{\Delta_m}{\delta_{in}})\mathcal{R}_m(\Delta_m,\omega) 
\rightarrow
\nonumber
\\
& M\int d\Delta_m  \tilde{G}_{r} (\frac{\Delta_m}{\delta_{in}}) \mathcal{R}(\Delta_m,\omega).
}
\label{eqfa-9}

Taking into account the continuous distribution of the frequencies of the miniresonators in the Eq. \eqref{a-omega}, after carrying out calculations we obtain

\fla{
\tilde{a}(\omega)=&
T_{cw}
(\omega)\tilde{A}_{in}(\omega)= 
\nonumber
\\
&\frac{\sqrt{\kappa}\tilde{A}_{in}(\omega)}{\left(\frac{1}{2}(\kappa+\gamma_c)+\frac{M g^2 }{{\delta_{in} \mathcal{F}_{1,2}(\chi\delta_{in},\chi \Gamma_{\Sigma},\omega)}} -i\omega \right) },
\label{a-omega-App}
}
where $\mathcal{F}_{1}(...)$ and $\mathcal{F}_{2}(...)$ correspond to two variants of resonator frequency distribution:

\fla{
\mathcal{F}_{1}(...)&= \frac{1}{\left(
\pi-[\phi_{(+)}(...)+\phi_{(-)}(...)]+i \ln 
[\mathcal{B}(...)]
\right)},
\label{F-1-AP}
\\
\mathcal{F}_{2}(...)&= \left(
1+\frac{\chi\Gamma_{\Sigma} -i\omega}{\chi\delta_{in}}
\right),
\label{F-2-AP}
}
where 
$\mathcal{B}(...)=\sqrt{\frac{(\frac{1}{2}\chi\delta_{in}+\omega)^2+(\chi\Gamma_{\Sigma})^2}{(\frac{1}{2}\chi\delta_{in}-\omega)^2+(\chi\Gamma_{\Sigma})^2}}$
and $\phi_{(\pm)}(...)=\arctan\left(\frac{2\chi\Gamma_{\Sigma}}{\chi\delta_{in}\pm 2\omega} \right)$ herewith  $\phi_{(\pm)}(...)<\pi/2$ if $\omega=0$.  

The analytical solution in equation \eqref{F-2-AP} can be very convenient for analyzing the properties of QM in the case of using narrow-band signal fields.
It is noteworthy that the analytical solution  \eqref{F-1-AP} was obtained with a realistic system of miniresonators.

\subsection{Dual CRIB-protocol}

First, we will focus on a general analysis of the possibility of implementing high efficiency in the studied QM scheme. 
Assuming a negligible effect of relaxation processes, we write a system of Heisenberg equations for the storage stage of the signal pulse.
Considering the fulfillment of the impedance matching condition  $ \hat{a}_{in}(t)=\sqrt{\kappa}\hat{a}(t)$, we get from Eqs.
\eqref{eqfa-1}-\eqref{eqfa-3}:

\fla{
\frac{\partial}{\partial t}  a & =  \frac{\kappa}{2} a - i \sum_{m}^{M} g_m b_{m},
\label{eqfa-1-n-S}
\\
 \frac{\partial b_{m} }{\partial t}& = - i\Delta_{m}  b_{m} 
-i g_m^{*} \hat{a} 
-i\sum_j  f_{j,m} S_{-}^{j,m},
\label{eqfa-2-n-S}
\\
 \frac{\partial S_{-}^{j,m} }{\partial t} & = -  i\delta_{j,m} S_{-}^{j,m}  
-i f_{j,m}^{*} b_m. 
\label{eqfa-3-n-S}
}

The completion of the stage of signal pulse storage is accompanied by the disappearance of the light field in the miniresonators, which, being absorbed by the atomic system, leads to the appearance of dephasing atomic coherence  $S_{-}^{j,m} (t)\sim S_{-}^{j,m} (t_r)e^{-i\delta_{j,m}t}$.
To restore the signal pulse, we inverts frequency detunings  of miniresonators and atoms at $t>t_r$ ($\Delta_{m}\rightarrow - \Delta_{m} $ and $\delta_{j,m}\rightarrow - \delta_{j,m} $ ). 
The emission of the photon echo $\hat{a}_e$ will be described by a system of equations:

\fla{
\frac{\partial}{\partial t}  a_e & =  -\frac{\kappa}{2} a_e - i \sum_{m}^{M} g_m b_{m},
\label{eqfa-1-e-S}
\\
 \frac{\partial b_{m} }{\partial t}& =  i\Delta_{m}  b_{m} 
-i g_m^{*} \hat{a}_e 
-i\sum_j  f_{j,m} S_{-}^{j,m},
\label{eqfa-2-e-S}
\\
 \frac{\partial S_{-}^{j,m} }{\partial t} & =  i\delta_{j,m} S_{-}^{j,m}  
-i f_{j,m}^{*} b_m. 
\label{eqfa-3-e-S}
}

The Eqs. \eqref{eqfa-1-e-S}-\eqref{eqfa-3-e-S}  describe a process that is inverse in time compared to the absorption of a signal pulse at the stage of its storage in quantum memory.
This follows from the fact that these equations coincide with Eqs. \eqref{eqfa-1-n-S}-\eqref{eqfa-3-n-S}  if time is reversed in Eqs. \eqref{eqfa-1-e-S}-\eqref{eqfa-3-e-S} ($t\rightarrow -t$) and  variables $a_e$ and $ S_{-}^{j,m}$ are redefined as follows: 
$a_e\rightarrow - a_e$, $S_{-}^{j,m}\rightarrow - S_{-}^{j,m}$).
The effects of atomic decoherence and resonator losses are discussed below (in subsection E) after examining the ROSE protocol.

\subsection{ROSE-protocol}

In this case, we use two  laser pulses, each of the has pulse area close to $\pi$ \cite{Damon2011}. These pulses provide rephasing of the atomic coherence created by the input signal pulse. 
This is one of the variants \cite{moiseev2011photon,Damon2011, mcauslan2011photon,ma2021elimination}, which  provides implementation of photon echo QM in atomic system with natural inhomogeneous broadening.

These pulses act on  atomic ensembles in M mini resonators.
To control the atomic rephasing, we can provide arbitrary phases $\phi_m$  of light pulses in each miniresonator.
Below, we will fix the optimal values of $\phi_m$.
Thus, the initial atomic coherence after exposure to two light pulses will be at $t=t_0$:

\fla{
 S_{-}^{j,m} (t_0)  &=
-i \sqrt{2\pi} f_{j,m}^{*} 
e^{ i\delta_{j,m} t_0}
\mathcal{Q}(t_0)
e^{i\phi_m}
\tilde{b}_{m,s}(\delta_{j,m}),
\label{atom-coherence-2-initial}
\\
\tilde{b}_{m,s}(\omega) & = - i g_m^{*} \frac{\chi T_{cw}(\omega)}{\left ( \chi\Gamma_{\Sigma}+i(\chi\Delta_{m}-\omega)\right)} 
\tilde{A}_{in}(\omega),
\label{b-m-omega-initial}
}
where $\mathcal{Q}(t_0)=\exp\{-T_s/T_{2,1\leftrightarrow2}\} \exp\{-(t_0-T_s) \gamma_a \}$ is a factor describing the influence of relaxation effects on atomic transitions $\ket{1}\leftrightarrow\ket{3}$ and 
$\ket{1}\leftrightarrow\ket{2}$ ($T_{2,1\leftrightarrow m}$ is a phase relaxation time on atomic transition $\ket{1}\leftrightarrow\ket{m}$),
$\tilde{b}_m(\delta_{j,m})$ is given by Eqs. \eqref{b-m-omega} and \eqref{a-omega}.

Here, taking into account the initial nonzero atomic coherence at time $t=t_0$ (see Fig.\ref{Long-lived_storage})), we get the following equations for the atomic coherence and the light field modes:

\fla{
\frac{\partial}{\partial t}  a & =  -\frac{1}{2}\left(\kappa+\gamma_c\right) a - i \sum_{m}^{M} g_m b_{m},
\label{eqfa-1-n}
\\
 \frac{\partial b_{m} }{\partial t}& = - \left(i\Delta_{m}+\gamma_b\right)  b_{m} 
\nonumber
\\
&-i g_m^{*} \hat{a} 
-i\sum_j  f_{j,m} S_{-}^{j,m},
\label{eqfa-2-n}
\\
 \frac{\partial S_{-}^{j,m} }{\partial t} & = - \left( i\delta_{j,m}+\gamma_a \right)  S_{-}^{j,m}  
-i f_{j,m}^{*} b_m
\nonumber
\\
&+\delta(t-t_0) S_{-}^{j,m} (t_0),
\label{eqfa-3-n}
}
where the initial coherence at the atomic transition $\ket{1}\leftrightarrow\ket{3}$  (at time $t=t_0$): 

\fla{
 S_{-}^{j,m} (t_0)  &=
e^{- i\delta_{j,m}(t_0-t_e)} e^{i\phi_m}
 \mathcal{D}_{-}^{j,m}(t_0),
\label{initial-coherence-2}
\\
\mathcal{D}_{-}^{j,m}(t_0)&=
-i \sqrt{2\pi} f_{j,m}^{*} 
\mathcal{Q}(t_0)
\tilde{b}_{m,s}(\delta_{j,m}).
}

Using the formal solution of Eq.\eqref{eqfa-3-n}

\fla{
 S_{-}^{j,m} (t>t_0)  &= 
 e^{- i\delta_{j,m}(t-t_e)-\gamma_a (t-t_0)}
 \mathcal{D}_{-}^{j,m}(t_0)
 \nonumber
 \\
 &-i f_{j,m}^{*}
 \int_{t_0}^{t}dt'
 e^{- i(\delta_{j,m}-i\gamma_a )(t-t')}b_m(t'),
\label{initial-coherence}
}
in Eq.\eqref{eqfa-2-n}, we obtain (see Eqs.\eqref{eqfa-5},\eqref{sum_int})

\fla{
& \frac{\partial {b}_{m} }{\partial t} =  - \left(i\Delta_{m}+\gamma_b\right)  \hat{b}_{m} 
-i g_m^{*} a_e
\nonumber
\\
&-\sqrt{2\pi} |f_m|^2 \mathcal{Q}(t_0) e^{i\phi_m} \sum_{j=1}^{N_m} 
 e^{- i\delta_{j,m}(t-t_e)-\gamma_a (t-t_0)}
 \tilde{b}_{m,s}(\delta_{j,m})
\nonumber
\\
& -  N_m |f_m|^2 
  \int_{t_e}^t dt' b_m (t')
  \exp\{-\left( \Delta_{in}+\gamma_a \right) (t-t')\},
 \label{eqfa-5-new}
}
taking into account the predominant value of the inhomogeneous broadening of the optical transition  $\ket{1}\leftrightarrow\ket{3}$, and the weak phase relaxation ($\mathcal{Q}(t)=\mathcal{Q}(t_e)$), we get:

\fla{
 &\frac{\partial {b}_{m} }{\partial t} =  - \left(i\Delta_{m}+\gamma_b+
 \Gamma_{a}^0 \right)  b_{m} 
-i g_m^{*} a_e
\nonumber
\\
&-\sqrt{2\pi} |f_m|^2 \mathcal{Q}(t_e) e^{i\phi_m} \sum_{j=1}^{N_m} 
 e^{- i\delta_{j,m}(t-t_e)}
 \tilde{b}_{m,s}(\delta_{j,m}).
 \label{eqfa-5-new2}
}
Substituting

\fla{
\tilde{b}_{m,s}(\omega)  =
- i \frac{g_m^{*}\chi}{\sqrt{\kappa}} \frac{1 }{\left ( \chi\Gamma_{\Sigma}+i(\chi\Delta_{m}-\omega)\right)} 
\tilde{A}_{in}(\omega),
}
in Eq.\eqref{eqfa-5-new2}, we find

\fla{
& \frac{\partial {b}_{m} }{\partial t} =  - \left(i\Delta_{m}+\gamma_b+
 \Gamma_{a}^0 \right)  b_{m} 
-i g_m^{*} a_e
\nonumber
\\
&+i\sqrt{2\pi} 
\Gamma_{a}^0 \mathcal{Q}(t_e) 
\frac{g_m^{*}\chi}{\pi\sqrt{\kappa}}
 e^{i\phi_m} \int d\omega 
  \frac{e^{- i\omega(t-t_e)} \tilde{A}_{in}(\omega) }{\left ( \chi\Gamma_{\Sigma}+i(\chi\Delta_{m}-\omega)\right)}.
  \label{eqfa-5-new2-1}
}
Turning to the Fourier picture 
(${b}_{m}(t)=\frac{1}{\sqrt{2\pi}}\int d\omega e^{-i\omega t} \tilde{b}_m(\omega)$, and ${a}_e(t)=\frac{1}{\sqrt{2\pi}}\int d\omega e^{-i\omega t} \tilde{a}_e(\omega)$), we obtain (additional prefactor $\sqrt{2\pi}$ in right side):

\fla{
&\left(i(\Delta_{m}-\omega)+\gamma_b+
 \Gamma_{a}^0 \right)  \tilde{b}_{m}(\omega) 
=-i g_m^{*} \tilde{a}_e(\omega)
\nonumber
\\
&+i g_m^{*}\mathcal{Q}(t_e) e^{ i\omega t_e}
e^{i\phi_m}
 \zeta(\Delta_m,\omega)   
\tilde{A}_{in}(\omega),
  \label{eqfa-5-new2-1}
}
where
\fla{
\zeta(\Delta_m,\omega)= 2 
    \frac{ \frac{\chi}{\sqrt{\kappa}}\Gamma_{a}^0 }{\left ( \chi\Gamma_{\Sigma}+i(\chi\Delta_{m}-\omega)\right)},
    \label{Zeta}
    }
and 

\fla{
& \tilde{b}_{m}(\omega) 
=
\frac{-i g_m^{*} \left[\tilde{a}_e(\omega)
-\mathcal{Q}(t_e) e^{ i(\omega t_e+\phi_m)} 
 \zeta(\Delta_m,\omega)   
\tilde{A}_{in}(\omega)\right]
}{\left(i(\Delta_{m}-\omega)+\gamma_b+
 \Gamma_{a}^0 \right) }.
  \label{eqfa-5-new2-1}
}

Using then  

\fla{
\left(\frac{\kappa+\gamma_c}{2}-i\omega \right) \tilde{a}_e(\omega) =
- i \sum_{m}^{M} g_m \tilde{b}_m(\omega),
\label{eqfa-1-Furier}
}
we find:

\fla{
&\left(\frac{\kappa+\gamma_c}{2}
+  \sum_{m}^{M} 
\frac{ |g|^2  }{\left(\gamma_b+
 \Gamma_{a}^0 +i(\Delta_{m}-\omega) \right)}
-i\omega \right) \tilde{a}_e(\omega) =
\nonumber
\\
&=  \mathcal{Q}(t_e) |g|^2   \sum_{m}^{M}
\frac{
 \zeta(\Delta_m,\omega) e^{ i(\omega t_e+\phi_m)}  
}{\left(\gamma_b+
 \Gamma_{a}^0+i(\Delta_{m}-\omega) \right) } \tilde{A}_{in}(\omega).
\label{a_echo-Furier}
}

\subsection{Lorentz distribution}

Making the transition from summation to integration over the frequencies of the miniresonators

\fla{
\sum_{m}^{M}...\rightarrow \int d \Delta_m \frac{\delta_{in}}{\pi(\delta_{in}^2+\Delta_m^2)}...
}

\textbf{A.Control of atomic coherence with constant phase $\phi_m=\phi_0$}.

By substituting  constant phase $\phi_m=\phi_0$ for each miniresonator, we obtain after calculating two integrals in Eq.\eqref{a_echo-Furier}:

\fla{
&\left(\frac{\kappa+\gamma_c}{2}
+\frac{ M |g|^2 }{\left(\gamma_b+
 \Gamma_{a}^0 +\delta_{in}-i\omega \right)}
-i\omega \right) \tilde{a}_e(\omega) =
\nonumber
\\
&
= \frac{ 2 M |g|^2  
\frac{\chi}{\sqrt{\kappa}}\Gamma_{a}^0}{\left ( \chi(\Gamma_{\Sigma}+\delta_{in})-i\omega)\right)}
\frac{\mathcal{Q}(t_e) e^{ i(\omega t_e+\phi_0)}
 \tilde{A}_{in}(\omega)
}{\left(\Gamma_{\Sigma}+\delta_{in} -i\omega\right) }.
\label{a_echo-2}
}
Now taking into account 
$\delta\omega_s\ll \delta_{in}+\Gamma_{\Sigma}$ and $\kappa  =  \gamma_c+
\frac{2M g^2 }{ \left(
\delta_{in}+\Gamma_{\Sigma}\right)}$, we get from Eq.\eqref{a_echo-2}:

\fla{
\tilde{a}_e(\omega) \cong
\mathcal{Q}(t_e)
 \frac{1}{\sqrt{\kappa}}
E_1 \frac{\Gamma_{a}^0 
}{\left( \Gamma_{\Sigma}+\delta_{in} \right) }
e^{ i(\omega t_e+\phi_0)}\tilde{A}_{in}(\omega).
\label{eqfa-1-Furier}
}
where $E_1=\frac{(\kappa-\gamma_c)}{\kappa}$ (see also comments to Eq.\eqref{storage}).

Finally taking into account the input-output relation:
$\tilde{A}_{echo}(\omega)=\tilde{A}_{out}(\omega)=
\sqrt{\kappa}\tilde{a}_e(\omega)
$, we find for the echo field:

\fla{
\tilde{A}_{echo}(\omega)
&\cong  E_{Lor}^{1/2}
e^{ i(\omega t_e+\phi_0)}\tilde{A}_{in}(\omega)
 ,
\label{eqfa-1-Furier}
\\
E_{Lor}^{1/2}&= \mathcal{Q}(t_e)
  E_1   \frac{\Gamma_{a}^0 
}{\left( \Gamma_{a}^0+\gamma_b+\delta_{in} \right) }.
\label{Eff_Lor_1}
}

Thus, in order to get high efficiency, it is necessary to have large enough coupling constant of atoms with miniresonator field $\Gamma_{a}^0\gg \delta_{in}$.

\textbf{B.Control of atomic coherence in $m$-th mini-resonance with different laser phases}. In this case, we assume $\phi_m=-2\arctan\{\frac{\Delta_m}{\Gamma_{\Sigma}} \}$ for the second laser pulse, which leads to:

\fla{
e^{-i\phi_m}\zeta(\Delta_m,0)= 2 
\frac{\Gamma_{a}^0 }{\sqrt{\kappa}\left ( \Gamma_{\Sigma}-i \Delta_{m}\right)},
\label{Zeta}
}

Taking again into account a sufficiently narrow spectrum of signal pulse  ($\delta\omega_s\ll \delta_{in}+\Gamma_{\Sigma}$), we obtain after integration over $\Delta_m$: 
\fla{
&\left(\frac{\kappa+\gamma_c}{2}
+\frac{ M |g|^2 }{\left(
 \Gamma_{\Sigma} +\delta_{in}\right)}
\right) \tilde{a}_e(\omega) =
\nonumber
\\
&
=2 M |g|^2  
\frac{\Gamma_{a}^0}{\sqrt{\kappa}}
\frac{\mathcal{Q}(t_e) e^{ i(\omega t_e+\phi_0)}
 \tilde{A}_{in}(\omega)
}{\Gamma_{\Sigma}\left(\Gamma_{\Sigma}+\delta_{in}\right) }.
\label{a_echo-2}
}

Then after simple calculation we find:

\fla{
& \tilde{A}_{out}(\omega)=\sqrt{\kappa} \tilde{a}_e(\omega) =
\nonumber
\\
&
=\mathcal{Q}(t_e) \frac{\Gamma_{a}^0}{(
 \Gamma_{a}^0+\gamma_b) }
\frac{(\kappa-\gamma_c)}{\kappa}
 e^{ i(\omega t_e+\phi_0)}
\tilde{A}_{in}(\omega).
\label{a_echo-2-perfect}
}

Thus, we see in Eq. \eqref{a_echo-2-perfect}, that  the input signal field can be emitted in a photonic echo with high efficiency if $\gamma_c\ll\kappa$, $\gamma_b\ll\Gamma_{a}$ and $\mathcal{Q}(t_e)\approx 1$.

\subsection{Rectangular frequency distribution}
Here we consider only the case of controlling the atomic coherence in the $m$-th miniresonator using laser pulses with added phases $\phi_m=-2\arctan\{\frac{\Delta_m}{\Gamma_{\Sigma}} \}$.   
Again, we replace  the summation in Eq. \eqref{a_echo-Furier} to the following  integration over the frequencies of the miniresonators:

\fla{
\sum_{m}^{M}...\rightarrow \frac{1}{\delta_{in}} \int_{-\delta_{in/2}}^{\delta_{in/2}} d \Delta_m ...
}

By performing similar calculations of integrals in Eq. \eqref{a_echo-Furier} 
we  take into account the relatively small spectral width.
After integration of Eq. \eqref{a_echo-Furier} over the miniresonator frequencies, we get

\fla{
&\left(\frac{\kappa+\gamma_c}{2}
+ \frac{2 M|g|^2}{\delta_{in}}
 \left[
 \frac{\pi}{2}-\arctan
 \left(\frac{2\Gamma_{\Sigma}}{\delta_{in}}
 \right)
 \right] \right) \tilde{a}_e(\omega) \cong
\nonumber
\\
&  2   
\frac{\mathcal{Q}(t_e)}{\sqrt{\kappa}}\frac{\Gamma_{a}^0}{\Gamma_{\Sigma}}   e^{ i\omega t_e}
\frac{2 M |g|^2}{ 
 \delta_{in} }
 \left[
 \frac{\pi}{2}-\arctan
 \left(\frac{2\Gamma_{\Sigma}}{\delta_{in}}
 \right)
 \right] 
 \tilde{A}_{in}(\omega).
\label{echo-rec-1}
}

Taking into account the  impedance matching  condition in further similar algebraic calculations, we again obtain a solution \eqref{a_echo-2-perfect} for the emitted photon echo field.

\subsection{Dual CRIB-protocol: decoherence effects}

Here we return to a more detailed study of the Dual CRIB-protocol, taking into account the effects of relaxation in the system of atoms and resonators.
We assume that at the retrieval stage we invert atomic and frequency detuning: $\delta_{j,m}\rightarrow -\delta_{j,m}$ and $\Delta_{m}\rightarrow -\Delta_{m}$ and
turning to the Fourier picture 
(${b}_{m}(t)=\frac{1}{\sqrt{2\pi}}\int d\omega e^{i\omega t} \tilde{b}_m(-\omega)$, and ${a}_e(t)=\frac{1}{\sqrt{2\pi}}\int d\omega e^{i\omega t} \tilde{a}_e(-\omega)$) and  the performed above calculations lead to the following equation:

\fla{
&\left(\frac{\kappa+\gamma_c}{2}
+  \sum_{m}^{M} 
\frac{ |g|^2  }{\left(\gamma_b+
 \Gamma_{a}^0 -i(\Delta_{m}-\omega) \right)}
+i\omega \right) \tilde{a}_e(-\omega) =
\nonumber
\\
&=  |g|^2  \sum_{m}^{M}  
\frac{\left[\mathcal{Q}(t_e) e^{ i\omega t_e}
 \zeta(\Delta_m,\omega)   
\tilde{A}_{in}(\omega)\right]
}{\left(\gamma_b+
 \Gamma_{a}^0-i(\Delta_{m}-\omega) \right) }.
\label{eqfa-1-Furier-nn}
}

By performing similar calculations transferring from the summation to the  integration in Eq.\eqref{eqfa-1-Furier-nn} in the case of a narrow spectrum of signal pulse $\delta\omega_s\ll \delta_{in}+\Gamma_{\Sigma}$,  
we obtain the following solution for the emitted echo signal:

\fla{
\tilde{A}_{echo,d-crib}(-\omega)
\cong E_1 E_2 \mathcal{Q}(t_e) 
 e^{ -i\omega t_e}\tilde{A}_{in}(\omega),
\label{echo-final}
}
where $E_2=\frac{\Gamma_a^0}{\Gamma_a^0+\gamma_b}$ (see also comments to Eq.\eqref{storage}).

In the time domain ($A_{echo,d-crib}(t)=\frac{1}{\sqrt{2\pi}}\int d\omega e^{i\omega t}\tilde{A}_{echo,d-crib}(-\omega)$), the echo pulse of Eq.\eqref{echo-final} efficiently retrieves a time-reversed copy of the input signal light pulse: 
${A}_{echo,d-crib}(t)\cong    \mathcal{Q}(t_e)  {A}_{in}[-(t-t_e)]$ (for $\gamma_b\ll \Gamma_{a}^0$,  $\gamma_c\ll\kappa$ and  $\mathcal{Q}(t_e)\approx 1$).

\subsection{Quantum noises}

Here we solve system of Eqs \eqref{eqfa-1-n}-\eqref{eqfa-3-n} but taking into account an arbitrary population of the atomic levels:

\fla{
\frac{\partial}{\partial t} \hat{a} & =  -\frac{\kappa}{2}\hat{a} - i \sum_{m}^{M} g_m \hat{b}_{m} +\sqrt{\kappa} \hat{a}_{in},
\label{eqfa-1-noise}
\\
 \frac{\partial\hat{b}_{m} }{\partial t}& = - i\Delta_{m} \hat{b}_{m} 
-i g_m^{*} \hat{a} 
-i\sum_j  f_{j,m} \hat{S}_{-}^{j,m},
\label{eqfa-2-noise}
\\
 \frac{\partial \hat{S}_{-}^{j,m} }{\partial t} & = - i\delta_{j,m}  \hat{S}_{-}^{j,m}  
+ i f_{j,m}^{*} \hat{b}_m \hat{w}_j+
\nonumber
\\
&+\delta(t-t_0) \hat{S}_{-}^{j,m} (t_0).
\label{eqfa-3-noise}
}
Here, for simplicity, we neglect  the influence of relaxation constants $\gamma_c$ and $\gamma_b$, keeping in mind the short evolution time.

Below we assume that at the initial moment of time $t=t_0$ there is a non-zero average difference in the populations of atoms $w(t_0)=\langle \hat{w}\rangle_{t_0}$ at the resonant transition 
$\hat{w}_j =\ket{3_j}\bra{3_j}-\ket{1_j}\bra{1_j}$
and, accordingly, an average probability of finding atoms at an excited level  $P_{33}^j=\langle \hat{P}_{33}^j\rangle_{t_0}$ (where $\hat{P}_{33}=\ket{3}\bra{3}$).

Replacing the operator  $\hat{w}_{j,m}$ by its average value over the ensemble of atoms  $w(t_0)$ and assuming that this average value changes negligibly over time, we find the following solutions for the operators following the calculation procedure used in the previous calculations

\fla{
 \hat{S}_{-}^{j,m} (t>t_0)  &= 
 e^{- i\delta_{j,m}(t-t_0)}
\hat{S}_{-}^{j,m} (t_0)+
 \nonumber
 \\
 & i f_{j,m}^{*} w(t_0)
 \int_{t_0}^{t}dt'
 e^{- i(\delta_{j,m}-i\gamma_a )(t-t')}\hat{b}_m(t').
\label{initial-coherence}
}

Substituting this solution into equation \eqref{eqfa-2-noise}, we find its solution, taking into account the large spectral width of the atomic transition

\fla{
\hat{b}_m(t)& = 
\hat{b}_m(t_0) e^{-i(\Delta_m+\Gamma_a(t_0))(t-t_0)}
\nonumber
\\
&-i g_m^{*} \int_{t_0}^t 
dt' e^{-i(\Delta_m+\Gamma_a(t_0))(t-t')}\cdot
\nonumber
\\
&\left(\hat{a} (t')  
-i\sum_j  f_{j,m} 
e^{- i\delta_{j,m}(t'-t_0)}
\hat{S}_{-}^{j,m} (t_0)
\right),
\label{b-m-noise}
}
where $\Gamma_a(t_0)=-w(t_0)\frac{N_a f_a^2}{\Delta_{in,a}}$.
It is worth noting that when atoms are at the lower level $w(t_0)\cong-1$, then  $\Gamma_a(t_0)=\Gamma_a^0$ 
and the atomic system absorbs radiation as it is in the quantum memory protocols studied. 

Under the condition $w(t_0)>0$, the atomic system already acts as an amplifier for the input radiation and the spontaneously emitted photons.
Considering that at the moment of time the modes min resonators are not excited, we can omit the term with the operator $\hat{b}_m(t_0)$ in the Eq. \eqref{b-m-noise}, as a term in the solution for the field mode operator $\hat{a}(t_0)$ and $\hat{a}_{in}$ of the general resonator In the absence of an external field in the waveguide.

Substituting this solution into Eq. \eqref{eqfa-1-noise}, we obtain the following solution for the general resonator mode operator

\fla{
\hat{a}(t) = &- fg^* \sum_m\sum_{j,m}
\int_{t_0}^t dt'
e^{-(\frac{\kappa}{2}+\ae(t_0))(t-t')}
\nonumber
\\
&\cdot\int_{t_0}^{t'}
d t" 
e^{-i(\Delta_m+\Gamma_a(t_0))(t'-t")}\cdot
\nonumber
\\
&\cdot e^{- i\delta_{j,m}(t''-t_0)}
\hat{S}_{-}^{j,m} (t_0),
\label{a-noise-photon}
}
where $\ae(t_0)=\frac{M|g|^2}{\delta_{in}+\Gamma_a(t_0)}$.

Using the solution \eqref{a-noise-photon}, we find the average number of photons spontaneously emitted into the common resonator $n_{r}(t)=\langle\hat{a}^{\dagger}(t)\hat{a}(t)\rangle$.
Taking into account in the calculation that at the beginning only resonant atoms are independently excited (miniresonators and the common resonator are empty), so that 
$\langle \hat{S}_{+}^{j,m} (t_0) \hat{S}_{-}^{j',m'} (t_ 0)\rangle=\delta_{j,m;j',m'}P^{j,m}_{33}(t_0)=P_{33}(t_0)>0$, we get:

\fla{
n_{r} (t)& = |fg|^2 P_{33}(t_0) 
\int_{t_0}^t dt' 
\int_{t_0}^{t'} d t"
e^{-(\frac{\kappa}{2}+\ae(t_0))(2t-t'-\tilde{t}')}
\nonumber
\\
&\cdot\int_{t_0}^{t}d \tilde{t}'
\int_{t_0}^{\tilde{t}' } d \tilde{t}^{''} 
e^{-\Gamma_a(t_0)[(t'-t")+(\tilde{t}'-\tilde{t}^{''})]} 
\nonumber
\\
& \cdot \sum_m\sum_{j,m} e^{-i\Delta_m[(t'-t")-(\tilde{t}'-\tilde{t}^{''})]}
e^{- i\delta_{j,m}(t''-\tilde{t}^{''})}.
\label{aa-noises}
}

In calculating the integrals in Eq. 
\eqref{aa-noises}, we first take into account the large inhomogeneous broadening of the atomic transition ($\Delta_{in,a}\gg\kappa,\delta_{in}$).
Then the response from such a group of atoms has a very small coherence time ($\sim \Delta_{in,a}^{-1}$), so that when calculating the integrals by time, we can replace
$\sum_{j,m}e^{- i\delta_{j,m}(t''-\tilde{t}^{''})}=
N_a\exp\{-\Delta_{in,a} (t''-\tilde{t}^{''}) \}\rightarrow 2 \rho_a\delta(t''-\tilde{t}^{''})$ (where $\rho_a=N_a/\Delta_{in,a}$ is a spectral density of atoms).
It is worth noting that taking into account the spectral width $\delta\omega_c$ of the exciting laser pulse acting on the resonant atoms will give the same result if $\delta\omega_c\gg\kappa $.
Using this relation we get after changing the integration limits ($\int_{t_0}^t dt' 
\int_{t_0}^{t'} d t"\rightarrow \int_{t_0}^t dt" 
\int_{t''}^{t} d t"$ and similarly for ($\tilde{t}'$, $\tilde{t}''$)):

\fla{
n_{r}(t)&= 2 \Gamma_a^0 |g|^2 P_{33}(t_0) \int_{t_0}^{t} dt" \int_{t''}^{t} dt' \int_{t''}^{t} d \tilde{t}'
 \cdot
\nonumber
\\
&
\cdot e^{-(\frac{\kappa}{2}+\ae(t_0))(2t-t'-\tilde{t}')}  
e^{-\Gamma_a(t_0)(t'+\tilde{t}'-2t")}
\nonumber
\\
& \cdot \sum_m e^{-i\Delta_m (t'-\tilde{t}')}.
\label{aa-noises-2}
}

In the text step, we use similar procedure for integration over the frequency detunings $\Delta_m$ 
$\sum_{m}e^{- i\Delta_{m}(t'-\tilde{t}^{'})}=
M\exp\{-\delta_{in} (t'-\tilde{t}^{'}) \}\rightarrow\frac{2 M}{\delta_{in}}\delta(t'-\tilde{t}^{'})$, which leads after integration over $\tilde{t}'$:

\fla{
n_{r}(t)&\cong 2 \Gamma_a^0 \kappa P_{33}(t_0) \int_{t_0}^{t} dt" \int_{t''}^{t} dt' \cdot
\nonumber
\\
& \cdot  
e^{-2(\frac{\kappa}{2}+\ae(t_0))(t-t')} e^{-2\Gamma_a(t_0)(t'-t")},
\label{aa-noises-2}
}
where we  have also taken into account the impedance matching condition (in the limit of large $\delta_{in}\gg\Gamma_a^0$)  ($\kappa  \cong \frac{2M g^2 }{ \delta_{in}}$).
Finally, calculating the double integral in Eq.\eqref{aa-noises-2}, we obtain:

\fla{
n_{r}(t)&\cong \frac{ \Gamma_a^0 \kappa}{\frac{\kappa}{2}+\ae(t_0)-\Gamma_a(t_0)} P_{33}(t_0)\cdot
\nonumber
\\
& \cdot  
\bigl\{\frac{1-e^{-2\Gamma_a(t_0)(t-t_0)}}{\Gamma_a(t_0)}-
\frac{1-e^{-2(\frac{\kappa}{2}+\ae(t_0))(t-t_0)}}
{\frac{\kappa}{2}+\ae(t_0)}
\bigr\}.
\label{aa-noises-3}
}
  
The analysis of Eq. \eqref{aa-noises-3} is given in the main part.

\bibliography{apssamp.bib} 

\end{document}